\documentclass[acmtog, nonacm]{acmart}

\usepackage{booktabs}
\usepackage{mathtools}
\usepackage{amsmath}
\usepackage{bm}
\usepackage{appendix}
\usepackage{enumitem}
\usepackage{float}
\usepackage{amsfonts}
\usepackage{multirow}
\usepackage{wrapfig}
\usepackage{graphicx}
\usepackage{listings}
\usepackage{array}
\usepackage{steinmetz}
\usepackage[final]{pdfpages}
\usepackage{xspace}
\usepackage{color}
\usepackage{xcolor}
\usepackage{setspace}
\usepackage{colortbl}
\usepackage[abbreviations]{glossaries-extra}
\usepackage[ruled]{algorithm2e}
\usepackage[
            type={CC},
            modifier={by-nc-sa},
            version={4.0},
            imageposition={left}
           ]{doclicense}

\newcommand\paraspace{-2.3mm}
\SetAlFnt{\small}
\SetAlCapFnt{\small}
\SetAlCapNameFnt{\small}
\SetAlCapHSkip{0pt}

\acmJournal{TOG}

\makeatletter
\def\@copyrightpermission{
\doclicenseThis
}
\makeatother

\AtBeginDocument{
  \providecommand\BibTeX{{
    \normalfont B\kern-0.5em{\scshape i\kern-0.25em b}\kern-0.8em\TeX}}}

\newcommand{\projectname}{\textit{HoloSpectra}\xspace}
\newcommand{\pIndex}{p}
\newcommand{\phs}{u}
\newcommand{\slmPhasePrimary}{\phs_\pIndex}
\newcommand{\optmSlmPhasePrimary}{\hat{\phs}_\pIndex}
\newcommand{\lossFunc}{\mathcal{L}}
\newcommand{\propKernel}{h_\pIndex}
\newcommand{\scale}{s}
\newcommand{\tgtIntensity}{I_\pIndex}
\newcommand{\numSubFrames}{T}
\newcommand{\subFrameIndex}{t}
\newcommand{\laserIntensity}{l}
\newcommand{\optmLaserIntensity}{\hat{\laserIntensity}}
\newcommand{\slmPhaseSubFrame}{\phs_\subFrameIndex}
\newcommand{\optmSlmPhaseSubFrame}{\hat{\phs}_\subFrameIndex}
\newcommand{\lossTerm}{L}
\newcommand{\wavelength}{\lambda}
\newcommand{\pAnchor}{\pIndex_{\text{anchor}}}
\newcommand{\slmPhaseMean}{\phs_\subFrameIndex^{\text{mean}}}
\newcommand{\slmPhaseOffset}{\phs_\subFrameIndex^{\text{offset}}}
\newcommand{\tv}{\nabla}
\newcommand{\imageLossThreshold}{\epsilon_{\text{image}}}
\newcommand{\optmScale}{\hat{\scale}}

\definecolor{Orange}{rgb}{1, 0.5, 0}
\definecolor{DarkGreen}{rgb}{0, 0.5, 0}
\definecolor{Purple}{rgb}{0.7, 0, 0.7}
\definecolor{Red}{rgb}{1.0, 0.0, 0.0}
\definecolor{DarkRed}{rgb}{0.5, 0.0, 0.0}
\definecolor{Brown}{rgb}{0.7, 0.4, 0.1}
\definecolor{Blue}{rgb}{0, 0, 1.}
\definecolor{DarkBlue}{rgb}{0, 0, 0.6}
\definecolor{Green}{rgb}{0., .6, 0.}
\definecolor{Custom}{rgb}{0.3, .1, 0.2}
\definecolor{Yellow}{rgb}{.9, .7, 0.}
\definecolor{Purple}{rgb}{.9, .1, 0.8}

\newcommand{\darkred}[1]{\textcolor{DarkRed}{#1}}
\newcommand{\darkblue}[1]{\textcolor{DarkBlue}{#1}}

\newabbreviation{cpd}{cpd}{Cycles Per Degree}
\newabbreviation{SLM}{SLM}{Spatial Light Modulator}
\newabbreviation{HVS}{HVS}{Human Visual System}
\newabbreviation{DMD}{DMD}{Digital Micromirror Device}
\newabbreviation{MLP}{MLP}{Multilayer Perceptron}
\newabbreviation{AR}{AR}{Augmented Reality}
\newabbreviation{VR}{VR}{Virtual Reality}
\newabbreviation{CGH}{CGH}{Computer-Generated Holography}
\newabbreviation{FoV}{FoV}{Field Of View}
\newabbreviation{HOE}{HOE}{Holographic Optical Element}
\newabbreviation{3D}{3D}{Three-Dimensional}
\newabbreviation{CNN}{CNN}{Convolutional Neural Network}
\newabbreviation{MTF}{MTF}{Modulation Transfer Function}
\newabbreviation{LC}{LC}{Liquid Crystal}
\newabbreviation{GD}{GD}{Gradient Descent}
\newabbreviation{DP}{DP}{Double Phase}
\newabbreviation{HDR}{HDR}{High Dynamic Range}
\newabbreviation{PSNR}{PSNR}{Peak Signal-to-noise Ratio}
\newabbreviation{SSIM}{SSIM}{Structural Similarity}
\newabbreviation{LPIPS}{LPIPS}{Perceptual Similarity Metric}

\global\long\def\LPIPS{\gls{LPIPS}\xspace}
\global\long\def\PSNR{\gls{PSNR}\xspace}
\global\long\def\SSIM{\gls{SSIM}\xspace}
\global\long\def\DP{\gls{DP}\xspace}
\global\long\def\GD{\gls{GD}\xspace}

\global\long\def\MLP{\gls{MLP}\xspace}
\global\long\def\HDR{\gls{HDR}\xspace}

\global\long\def\HVS{\gls{HVS}\xspace}
\global\long\def\3D{\gls{3D}\xspace}

\global\long\def\SLM{\gls{SLM}\xspace}
\global\long\def\AR{\gls{AR}\xspace}
\global\long\def\VR{\gls{VR}\xspace}

\global\long\def\CGH{\gls{CGH}\xspace}

\newcommand{\etal}{~et al.\@\xspace}

\newcommand{\eg}{e.g.\@\xspace}

\newcommand{\ie}{i.e.\@\xspace}

\newcommand{\refSec}[1]{Sec.~\ref{sec:#1}}
\newcommand{\refFig}[1]{Fig.~\ref{fig:#1}}
\newcommand{\refFigFull}[1]{Figure~\ref{fig:#1}}
\newcommand{\refEq}[1]{Eq.~(\ref{eq:#1})}
\newcommand{\refTbl}[1]{Tbl.~\ref{tbl:#1}}

\copyrightyear{2023}
\setcopyright{cc}
\begin{document}

\title{Multi-color Holograms Improve Brightness in Holographic Displays}

\author{Koray Kavaklı}
\affiliation{
 \institution{Koç University}
 \country{Türkiye}
}
\email{kavakli@ku.edu.tr}

\author{Liang Shi}
\authornote{denotes corresponding authors}
\affiliation{
 \institution{Massachusetts Institute of Technology}
 \country{United States of America}
}
\email{liang@mit.edu}

\author{Hakan Urey}
\affiliation{
 \institution{Koç University}
 \country{Türkiye}
}
\email{hurey@ku.edu.tr}

\author{Wojciech Matusik}
\affiliation{
 \institution{Massachusetts Institute of Technology}
 \country{United States of America}
}
 \email{wojciech@mit.edu}

\author{Kaan Akşit}
\authornotemark[1]
 \affiliation{
 \institution{University College London}
 \country{United Kingdom}
}
\email{k.aksit@ucl.ac.uk}

\begin{abstract}
Holographic displays generate Three-Dimensional (3D) images by displaying single-color holograms time-sequentially, each lit by a single-color light source.
However, representing each color one by one limits brightness in holographic displays.
This paper introduces a new driving scheme for realizing brighter images in holographic displays.
Unlike the conventional driving scheme, our method utilizes three light sources to illuminate each displayed hologram simultaneously at various intensity levels.
In this way, our method reconstructs a multiplanar three-dimensional target scene using consecutive multi-color holograms and persistence of vision.
We co-optimize multi-color holograms and required intensity levels from each light source using a gradient descent-based optimizer with a combination of application-specific loss terms.
We experimentally demonstrate that our method can increase the intensity levels in holographic displays up to three times, reaching a broader range and unlocking new potentials for perceptual realism in holographic displays.
\end{abstract}


\begin{teaserfigure}
\includegraphics[width=\textwidth]{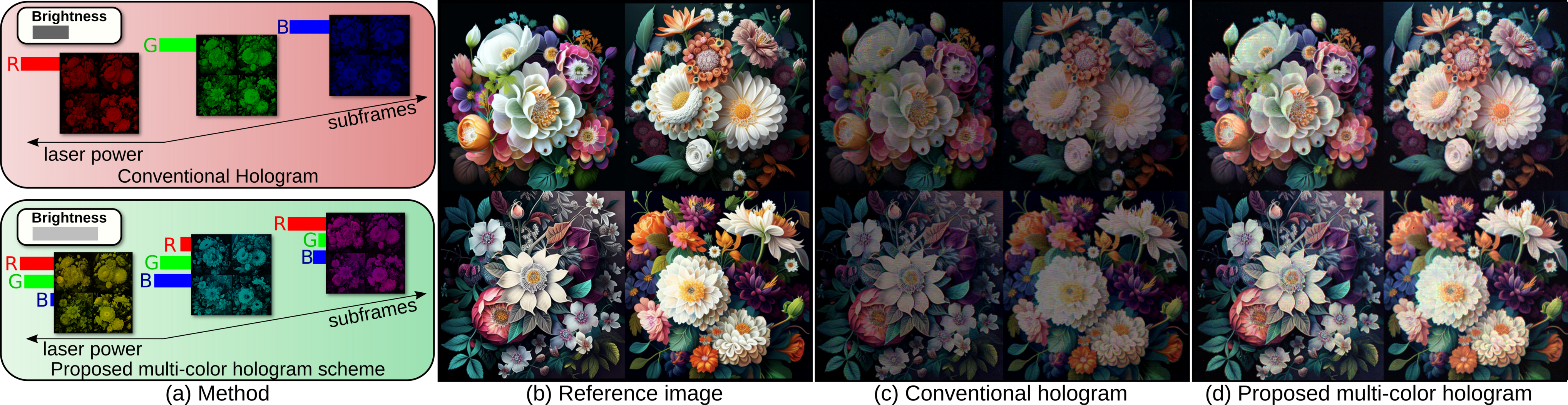}
\caption{
Our framework simultaneously uses multiple laser light sources to support brighter images in holographic displays.
(a) Conventional holograms display full-color images using single-color holograms, each dedicated to a color channel and illuminated by a single laser light source.
Our method instead optimizes multi-color holograms, each lit by and modulates multiple laser light sources.
Given a reference image (b),
photographs captured from a holographic display prototype with an 80 ms exposure:
(c) A conventional hologram reconstructs an image with limited brightness, and
(d) A multi-color hologram reconstructs a brighter image
(Source image: Midjourney, Link: \href{https://github.com/complight/images}{\textbf{Github:complight/images}}).
}
\label{fig:teaser}
\end{teaserfigure}

\maketitle
\section{Introduction}
\label{sec:introduction}
\begin{wrapfigure}{l}{0.27\columnwidth}
\centering
\vspace{-4mm}
\hspace{6mm}
\includegraphics[width=1.2\linewidth]{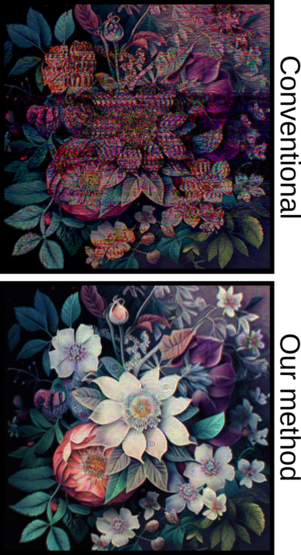}
\vspace{-6mm}
\caption{
Photographs showing conventional \textbf{(top)} and our \textbf{(bottom)} results when targeting $\times1.8$ brightness (Source image: Midjourney, 100 ms exposure).
}
\label{fig:distortions}
\vspace{-4mm}
\end{wrapfigure}

Recent advances in holographic displays~\cite{koulieris2019near} offer unique opportunities, such as the generation of high-quality \3D images at interactive rates~\cite{shi2022end} and slim eyeglasses-like form factors for \AR glasses~\cite{jang2022waveguide} and \VR headsets~\cite{kim2022holographic}.
However, holographic displays have yet to prove themselves in achieving perceptual realism, and one of the roadblocks is their brightness levels.
Conventional holographic displays use a single \SLM and reconstruct full-color images by time-sequentially displaying single-color holograms, each dedicated to a color channel~\cite{Pi22color}.
When holographic displays reconstruct scenes with intensity (brightness) levels beyond the light source peak intensity of their corresponding color channels, the result could often lead to darker images than the intended levels and produce visual distortions or color mismatches~(see \refFig{distortions} top).
In such cases, the brightness range of the target is typically limited to the peak intensity of the light source (see \refFig{teaser}(c)), which is often not enough to deliver the desired visual experience.
Alternatively, these displays could adopt light sources with higher power ratings.
However, high-power light sources pose an eye safety risk for users, create undesired heat, and increase hardware cost\footnote{Thorlabs HL6322G 15mW laser diode (\$77.45) is three times the price of Thorlabs HL6312G 5mW (\$24.45) as of August 8th, 2023.} and complexity (e.g., more powerful cooling unit), specifically for mobile or wearable display applications.
Thus, we are left with the question, \textit{``Can holographic displays better utilize their existing hardware resources to improve their brightness?''}
Without altering hardware, we argue that holographic displays could dedicate extra time to each color channel to improve their perceived intensity levels, as demonstrated in \refFig{teaser}(d).
Our work aims to improve holographic displays' brightness more effectively by aggressively utilizing color primaries and holograms.
For this purpose, we introduce a new \CGH driving scheme using multi-color holograms.
In this scheme, multi-color holograms simultaneously operate over multiple wavelengths of light and provide \3D multiplanar images.
We calculate multi-color holograms using a Gradient Descent (GD) based solver guided by a combination of application-specific loss functions.
In the meantime, we co-optimize the intensity levels required to illuminate each multi-color hologram.
We experimentally verify our findings using a holographic display prototype by showing reconstructions of brighter scenes artifact-free and color-accurate manner.
Specifically, our work (\href{https://github.com/complight/multicolor}{\textbf{GitHub:complight/multicolor}}) introduces the following contributions:

\begin{itemize}
\item Multi-Color Hologram Driving Scheme.
A new \CGH scheme that co-optimizes multi-color holograms and laser powers for each subframe using a GD-based solver with a combination of application-specific loss functions, leading to brighter images.

\item Experimental Verification.
We demonstrate artifact-free and color-accurate experimental results on a holographic display with a 1080p SLM driven by our multi-color hologram scheme.
We show a machine-learning model representing color production in our hardware can help guarantee color accuracy in image generation.
\end{itemize}
\vspace{\paraspace}

\section{Related Work}
\label{sec:related_work}
We survey the literature on multi-color holograms, dynamic ranges, brightness, and color production in holographic displays.
Beyond our survey, readers can consult to \CGH review by Chang\etal~\shortcite{Chang:20}.
\vspace{\paraspace}

\subsection{Brightness in Conventional Displays}
We define brightness as the highest intensity achievable by a display and dynamic range as the ratio between the highest and lowest intensity values.
Supporting \HDR in conventional displays has been under development for over two decades~\cite{seetzen2004high}.
Today's conventional \HDR display products offer smartphone-like intensity levels while their research counterparts could offer cloudy sky-alike intensity levels~\cite{zhong2021reproducing}.
There are also emerging research variants for \HDR \VR displays~\cite{matsuda2022realistic}.
In parallel, researchers investigate improving color production in a display using either a fewer~\cite{huang2017mixed} or larger~\cite{kauvar2015adaptive} number of color primaries.
Concerning conventional displays, holographic displays promise to generate a larger color gamut using coherent sources while promising a broader dynamic range and brightness~\cite{damberg2016high}.
Our work resembles an attempt to understand how much of this promise could be fulfilled in holographic displays more effectively with multi-color holograms.
\vspace{\paraspace}

\paragraph{Hardware approaches.}
The pixel depth of a phase-only \SLM~\cite{lee2009emerging} used in a holographic display dictates the color production accuracy of reconstructed images.
Although there are works improving brightness and color accuracy in SLMs~\cite{albero2013second,perez2016first, davis2020spatial}, these works aim to function as beam-shaping devices but generate images like an actual display would show.
A newly emerging technology, piston-mode-based phase modulators~\cite{oden2020innovations}, can offer four-bit quantization in phase for holographic display applications~\cite{choi2022time}.
An active research topic, nanophotonic phase arrays are also being investigated as a new type of \SLM for holographic displays~\cite{jabbireddy2022sparse}.
Our multi-color hologram driving scheme can be helpful for various SLMs.
But each new \SLM type would lead to a specific but not an SLM-universal solution.
Thus, we limit the discussion to LC-based phase-only SLMs, the most common type used in holographic displays.
\vspace{\paraspace}

\paragraph{Software approaches.}
Previous works capture images from holographic displays using \HDR imaging to improve the image quality algorithmically~\cite{yonesaka2016high,lee2015experimental}.
The work by Kadis\etal~\shortcite{kadis2022high} explored the performance of hologram optimizations concerning the bit-depth of a target image.
Chao\etal~\shortcite{chao2023brightness} proposed a light-efficiency loss function to enhance brightness.
Our work also tackles improving brightness in holographic displays.
\vspace{\paraspace}

\subsection{Multi-color Holograms for Holographic Displays}
\label{subsec:multi_color_holograms_in_holographic_displays}
Almost all hologram types, including rainbow holograms~\cite{choo2018image} or conventional Holographic Optical Elements (HOEs)~\cite{jang2020design} could be illuminated by a broadband light source.
However, illuminating these holograms leads to reconstructions of distorted or spatially-separated images.
To our knowledge, having such holograms be designed or optimized to operate simultaneously with multiple wavelengths of light is a rarity unless these holograms serve as a fixed-function optical component for beam-shaping or steering~\cite{cakmakci2021holographic} (\eg relay lens, mirror, or similar).
Fourier Rainbow holograms with incoherent light sources~\cite{yang2019full, kozacki2018fourier} help map the same image to a different perspective (directions) in the Fourier plane.
Yolalmaz and Yüce~\cite{yolalmaz2022comprehensive} introduce a deep-learning model that could generate holograms at various depths using multiple colors.
Previous works did not involve improving brightness levels by optimizing multi-color holograms and their light dosages.
\vspace{\paraspace}

\section{Multi-color Hologram Driving Scheme}
\label{sec:method}
%
\paragraph{Synthesizing Conventional Holograms.}
Existing holographic displays use the field-sequential color method, which replays three single-primary images (R, G, B) in rapid succession and relies on the \HVS to fuse them into a full-color image~\cite{Pi2022review}.
At any given time, only one monochromatic light source operates in the field-sequential method. 
Thus, a phase pattern is independently identified explicitly for this active wavelength.
For a full-color image, a conventional hologram is composed of three single-color phase patterns for each color primary and is subject to resolving the following optimization problem,
\begin{equation}
\optmSlmPhasePrimary \leftarrow \operatorname*{argmin}_{\slmPhasePrimary} \sum_{\pIndex=1}^{3} \lossFunc({\lvert e^{i\slmPhasePrimary} * \propKernel \rvert}^2, \scale \tgtIntensity),
\label{eq:conventional_optm}
\end{equation}
where $\pIndex$ denotes the index of a color primary, $\slmPhasePrimary$ is the SLM phase (for the active primary, abbreviated thereafter), $\optmSlmPhasePrimary$ is the optimized SLM phase, $\propKernel$ is the wavelength-dependent light transport kernel~\cite{matsushima2009band, kavakli2022learned}, $\tgtIntensity$ is the target image intensity, $\scale$ is an intensity scaling factor, set by default to 1, $*$ denotes the convolution operation, and $\lossFunc$ denotes any proper loss function that measures the difference between the reconstruction and target.
In \refEq{conventional_optm}, the SLM phase $\slmPhasePrimary$ is a 2D matrix with values ranging between -$\pi$ and $\pi$.
It can be encoded from a complex field through \DP method~\cite{Maimone2017-sg, shi2021towards}.
Recent works have demonstrated that coupling \DP with \GD optimizations can improve image quality~\cite{kavakli2023realisticdefocus}.
We use the same strategy in our optimizations, and their approach of coupling \DP with \GD optimizations~\cite{kavakli2023realisticdefocus} refers to the conventional method in the rest of this manuscript.
In a conventional hologram, setting $\scale$ such that the total intensity of the scaled image is beyond the intensity output of the single-primary sub-frame makes it challenging to produce distortion-free images (see \refFig{distortions}(top)).
This challenge can be more prominent when the propagation distance is short, as smaller sub-holograms are used to produce high peak intensities in a final image.
Thus, this challenge formulates the base of the problem we tackle in this work.
\vspace{\paraspace}

\begin{figure*}[hbt!]
\centering
\includegraphics[width=0.95\linewidth]{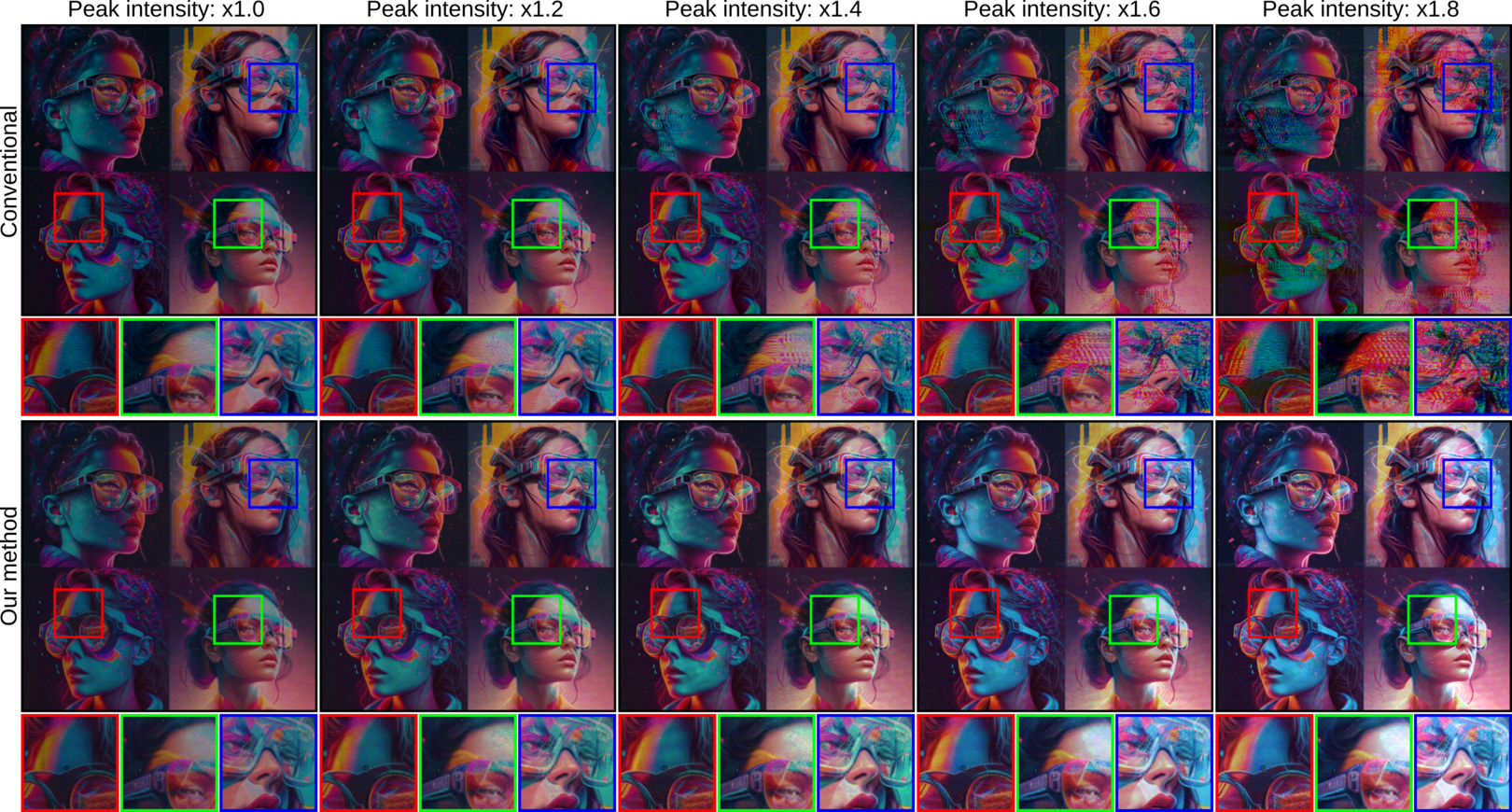}
\vspace{-1mm}
\caption{
Increasing peak intensity levels with our multi-color hologram scheme.
Photographs show that our method can enhance the peak intensity levels of the captures up to $\times1.8$ without noticeable artifacts or distortions.
In contrast, the conventional hologram fails to support beyond $\times1.0$ (Source image: Midjourney, Link: \href{https://github.com/complight/images}{\textbf{Github:complight/images}}, 140 ms exposure).
}
\label{fig:brightness_increase}
\vspace{-4mm}
\end{figure*}

\paragraph{Synthesizing Multi-Color Holograms.} 
Our solution to improve brightness and color production in holographic displays and requires a power-tunable light source -- often readily available in consumer laser light engines.
Our multi-color hologram scheme typically involves optimizing three-phase patterns, each illuminated by multiple color primaries with various light dosages, and a multi-color hologram combines these multi-color phase patterns.
Let $\numSubFrames$ be the total subframes for reproducing one color image (\ie, 3 in the case of conventional holograms).
Note that this is not to be confused with the repetition of subframes in time-multiplexing holography, which aims to reduce speckle noise~\cite{Lee2022-tz, choi2022time}.
Our method formulates the optimization problem as
\begin{equation}
\optmSlmPhaseSubFrame, \optmLaserIntensity_{(\pIndex, \subFrameIndex)} 
\leftarrow
\operatorname*{argmin}_{\slmPhaseSubFrame, \laserIntensity_{(\pIndex, \subFrameIndex)}} \underbrace{\sum_{\pIndex=1}^{3} \left\lVert \left(\sum_{\subFrameIndex=1}^{\numSubFrames} \left\lvert \laserIntensity_{(\pIndex, \subFrameIndex)} e^{i\frac{\wavelength_{\pIndex}}{\wavelength_{\pAnchor}}\slmPhaseSubFrame} * \propKernel \right\rvert^2\right) - \scale \tgtIntensity\right\rVert_2^2}_{\lossTerm_{\text{image}}},
\label{eq:holohdr_optm}
\end{equation}
where $\laserIntensity_{(\subFrameIndex, \pIndex)}$ represents the laser amplitude for the $\pIndex$-th primary at the $\subFrameIndex$-th subframe, $\wavelength_\pIndex$ denotes the wavelength of the active primary, $\wavelength_{\pAnchor}$ denotes the wavelength of the anchor primary, for which the nominal value of the SLM phase is calibrated against (\eg $\wavelength_{\pAnchor} = 515~nm$ in our hardware prototype). 
When $\numSubFrames = 2$ or $\numSubFrames = 1$, our method can operate at a higher fresh rate.
Note that  $\numSubFrames = 3$ offers better color accuracy over fewer subframes.
To speed up convergence and improve experimental results, our method extends optimizations with two additional losses in practice for a robust multi-color hologram generation,
\begin{equation}
\lossTerm_{\text{total}} = w_1\lossTerm_{\text{image}} + w_2\lossTerm_{\text{laser}} + w_3\lossTerm_{\text{variation}}.
\label{eq:loss_total}
\end{equation}
Here, $w_1,w_2,w_3$ are weights of each loss ($w_1=3.0, w_2=0.05, w_3=0.1$ in our implementation).
The laser loss $\lossTerm_{\text{laser}}$ is given by
\begin{equation}
\lossTerm_{\text{laser}} = \sum_{\pIndex=1}^{3} {\left ( \left(\sum_{\subFrameIndex=1}^{\numSubFrames} \laserIntensity_{(\pIndex, \subFrameIndex)}^2\right) - \text{max}(\tgtIntensity)\scale  \right)}^2.
\label{eq:laser_loss}
\end{equation}
For every color primary, $\lossTerm_{\text{laser}}$ encourages the sum of laser intensities across the subframes to match the scaled maximum intensity of the target image.
It accelerates the convergence of $\lossTerm_{\text{image}}$ and consistently produces more accurate color in complex scenes (see \refSec{evaluation} for an ablation study).
Depending on a targeted scene, there are the risks of laser powers at some subframes getting stuck at zero power or utilized less evenly.
To avoid such risks, we augment $\lossTerm_{\text{laser}}$ with a few additional terms described in the supplementary.
The variation loss $\lossTerm_{\text{variation}}$ is given by
\begin{multline}
\lossTerm_{\text{variation}} = \sum_{\subFrameIndex=1}^{\numSubFrames} \Biggl( \left\lVert \tv\left(\slmPhaseMean+\slmPhaseOffset\right) \right\rVert_2^2 + \left\lVert \tv\left(\slmPhaseMean-\slmPhaseOffset\right) \right\rVert_2^2 + \\
+\sigma\left(\slmPhaseMean+\slmPhaseOffset\right) + \sigma\left(\slmPhaseMean-\slmPhaseOffset\right) \Biggl),
\label{eq:variation_loss}
\end{multline}
where $\tv$ denotes the total variation operator, $\sigma(\cdot)$ denotes the standard deviation operator,
\begin{equation}
\slmPhaseSubFrame(x,y) = \begin{dcases}
    \slmPhaseMean(x,y)+\slmPhaseOffset(x,y), & x+y \ \text{is odd}\\
    \slmPhaseMean(x,y)-\slmPhaseOffset(x,y), & x+y \ \text{is even}\\
\end{dcases}.
\label{eq:db_phase}
\end{equation}

In our implementations, we use the total variation loss over an image pyramid of the reconstructed images.
Here, we use a variant of the traditional double-phase formula to obtain the solution.
Specifically, we add or subtract an offset phase $\slmPhaseOffset$ from a mean phase $\slmPhaseMean$ to obtain a low phase and a high phase for double phase interlacing (\refEq{db_phase}).
The variation loss discourages rapid change and large standard deviation for the low and high phase maps.
It reduces the speckle artifacts commonly appearing in the experiments and accelerates the convergence of $\lossTerm_{\text{image}}$.
\vspace{\paraspace}

\paragraph{Multi-color holograms with dynamic intensity scale.}
When manually setting $\scale$ close to its theoretical limit (3 in case of $\numSubFrames$=3), a high-quality reproduction is not always guaranteed.
Instead of finding the highest $\scale$ through trials or always using a low $\scale$ attainable for almost all scenes, we can jointly optimize $\scale$ to be as high as possible under a user-specified image loss threshold $\imageLossThreshold$,
\begin{align}
\optmSlmPhaseSubFrame, \optmLaserIntensity_{(\pIndex, \subFrameIndex)}, \optmScale \leftarrow \operatorname*{argmin}_{\slmPhaseSubFrame, \laserIntensity_{(\pIndex, \subFrameIndex)}, \scale} \lossTerm_{\text{total}} - w_4\scale \ , \ \ & \text{if} \ \ \lossTerm_{\text{image}} < \imageLossThreshold \\
\optmSlmPhaseSubFrame, \optmLaserIntensity_{(\pIndex, \subFrameIndex)} \leftarrow \operatorname*{argmin}_{\slmPhaseSubFrame, \laserIntensity_{(\pIndex, \subFrameIndex)}} \lossTerm_{\text{total}} \ , \ \  & \text{if} \ \ \lossTerm_{\text{image}} \geq \imageLossThreshold,
\label{eq:dynamic_scale}
\end{align}
where $w_4$ is the weight hyperparameter for the scale.
In \refSec{evaluation}, we show how this conditional update strategy helps discover a content-dependent maximum scale.
\vspace{\paraspace}

\begin{figure}[hbt!]
\centering
\includegraphics[width=0.93\linewidth]{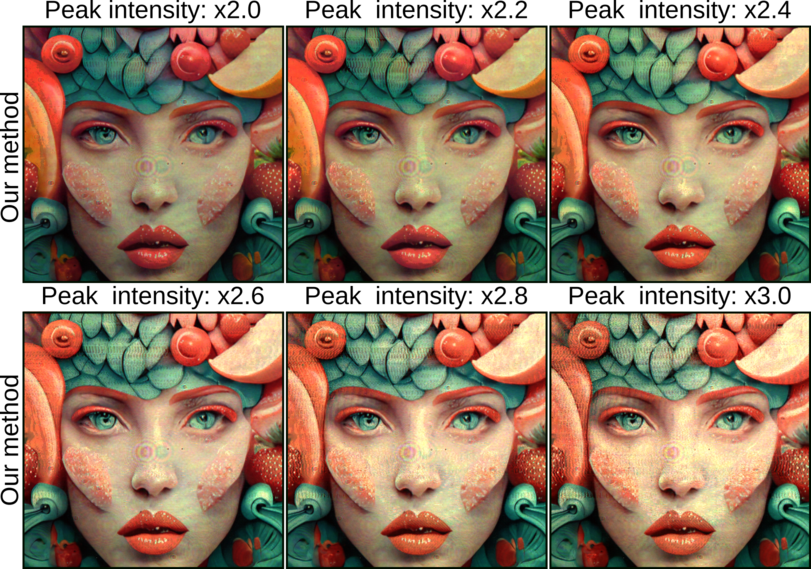}
\caption{
Photographs showing our method generating higher brightness beyond $\times 2.0$ (Source image: Midjourney, Link: \href{https://github.com/complight/images}{\textbf{Github:complight/images}}, 50 ms exposure).
}
\label{fig:beyond_18}
\vspace{-6mm}
\end{figure}

\section{Evaluation}
\label{sec:evaluation}
This section evaluates our method in terms of the achieved image brightness and color production. 
It also provides an ablation study to identify the contribution of each regularization term in \refSec{method}.
All our assessments are camera-captured from a holographic display prototype using three subframes, $\numSubFrames = 3$ (unless indicated otherwise).
Our prototype uses a Ximea MC245CG-SY camera to capture results and a Holoeye Pluto-VIS \SLM to display results.
Readers can consult the supplementary for more details of the display prototype.
\vspace{\paraspace}

\paragraph{Brightness.} 
\refFigFull{brightness_increase} shows photographs from our holographic display for conventional and multi-color schemes (more sample results in \refFig{extra_results} and supplementary).
For such a scene in \refFigFull{brightness_increase}, our scheme can safely support up to $\times1.8$ peak intensity without causing significant image distortions or artifacts.
On the other hand, the conventional hologram fails to support peak intensities higher than $\times1.0$ as in \refFig{brightness_increase} and \refFig{distortions}.
Beyond $\times1.8$ peak intensity levels, images are typically heavily dominated by noise in the conventional case.
In contrast, our case loses color integrity slightly or generates noises similar to the $\times1.2$ conventional case (see \refFig{beyond_18}).
\vspace{\paraspace}

\paragraph{Power rating.}
For intensities beyond $\times1.0$, the conventional holograms demand optical power ratings beyond $\times1.0$ to match the brightness levels.
The datasheet of a sample class 3B laser (Thorlabs HL6321G) reveals that the electrical input power ratings as 180~mW, 200~mW, and 220~mW for $\times1.0$ (5~mW optical power), $\times2.0$, and $\times3.0$ peak intensities, respectively. 
On the other hand, our method could satisfy the same brightness by running at maximum $\times1.0$ peak intensity in the worst case demanding the input electrical power of $140~mW$ with a class 3R laser (Thorlabs HL6312G/13G). 
In this design example, our methods IEC Class 3R lasers pose a low risk, while IEC Class 3B's direct exposure could induce retinal and skin injury~\cite{schulmeister2010risk}. 
Specifically, battery-operated wearable displays could relax their power rating and cost for component selection while users experience enhanced brightness levels with lower risks. 
We use the same laser and power rating in our assessments to compare both methods fairly. But our method uses a longer turn-on time for achieving brighter images.
\vspace{\paraspace}

\paragraph{Multi-Color Dynamic Intensity Scaling.}
Supporting an artifact and distortion-free solution strictly at $\times1.8$ peak intensity levels is not always guaranteed with our method, as each target scene's content heavily influences the results.
Therefore, we also offer a dynamic scale option as introduced in \refSec{method}.
\begin{wrapfigure}{l}{0.27\columnwidth}
\centering
\hspace{6mm}
\vspace{-6mm}
\includegraphics[width=1.2\linewidth]{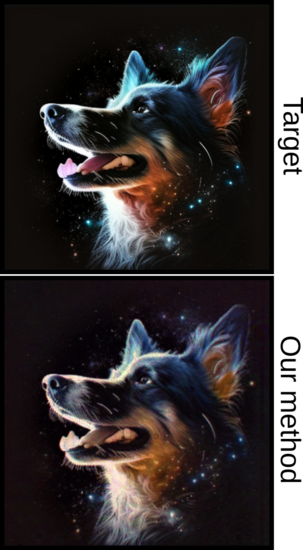}
\vspace{-2mm}
\caption{
Multi-color dynamic intensity scales to $\times 1.63$ brightness (Source image: Midjourney, 100 ms exposure).
}
\vspace{-5mm}
\label{fig:dynamic_scaling}
\end{wrapfigure}
\refFigFull{dynamic_scaling} shows a sample result from this dynamic intensity scale approach when enforcing the image loss to stay below a fixed value (0.01).
In this sample result, the dynamic intensity scale for our method automatically chooses the intensity level of a targetted scene as  $\times1.63$ rather than hardcoding as any other value (\eg $\times1.8$).
Thus, the dynamic intensity scale for our method offers a content-adaptive solution for choosing peak intensity levels.
For both conventional and multi-color cases, we measure the optical power using a Thorlabs PM100D power meter console equipped with Thorlabs S120VC and a calibrated camera from Radiant Imaging for intensity measurements.
In our next, we envision applying our method to \HDR targets so that an explicit definition of scale is no longer needed. 
We plan to tone map to a specific dynamic range for consistent brightness across across-frames in moving images. 
However, we clarify that this needs to be a thoroughly investigated in the future. 
\vspace{\paraspace}

\begin{figure}[hbt!]
\centering
\includegraphics[width=0.93\columnwidth]{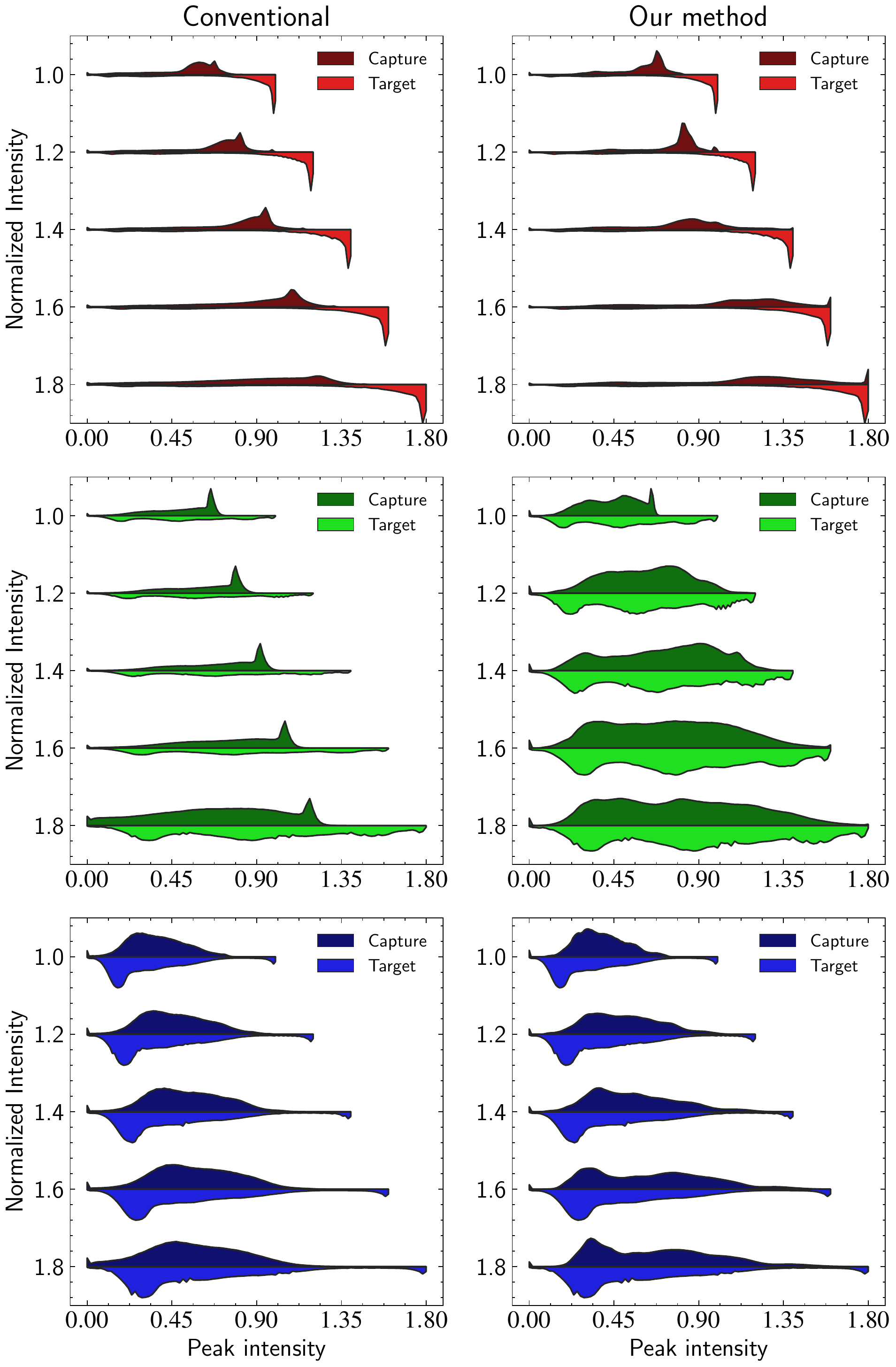}
\vspace{-2mm}
\caption{
Comparing red, green, and blue histograms of a target image with conventional and multi-color schemes for varying intensity levels (Same target as \refFig{extra_results}, $\times 1 - \times1.8$ intensity, 140 ms - 240 ms exposure).
}
\label{fig:color_analysis}
\vspace{-6mm}
\end{figure}

\paragraph{Controlling lasers.}
Our multi-color optimization routine provides normalized laser power estimates between one and zero.
This range is in the arbitrary unit and does not correspond to a physical value.
Thus, we must find a way to convert these normalized laser power estimates into meaningful values for our laser drivers.
For this purpose, we capture photographs from our prototype with various brightness values.
We separate the pixel levels for each photograph in the dataset for our photographs' red, green, and blue channels.
We normalize these sums and are left with the laser power settings we provided to capture the normalized sums (note that this assumption, we treat our camera's response as a linear response --relation between power and pixel levels.)
We use the laser settings and normalized sums to train a four-layer multilayer perceptron, where input is a normalized sum value, and output is the laser driver setting.
We provide the estimated laser power value from our optimizer to our learned model to get the laser power settings for our actual holographic display prototype.
\vspace{\paraspace}

\paragraph{Color production.} 
Accurately reproducing colors for a scene can be complex since it also involves identifying the relationship between laser control and image intensity.
As illustrated in previous figures, there is already a visible increase in intensity in multi-color holograms.
We must, however, assess whether these results are faithful reproductions of the target scene's color.
To improve color reproduction, we built a \MLP model to control the colors generated by our method.
Specifically, this \MLP with four hidden layers identifies the relationship between the laser powers suggested by the optimization, $\optmLaserIntensity_{(\pIndex, \subFrameIndex)}$, and values provided to the laser driver (see supplementary).
We evaluate the color reproduction of our results in \refFig{color_analysis} by comparing the color histogram of a target scene, the conventional hologram reconstruction, and the multi-color hologram reconstruction for each color primary.
Our method's histogram approximates the target, whereas the conventional hologram fails to follow the trend beyond $\times 1.0$ peak intensity.
We underline that our method does not aim for color gamut enhancement.
In addition, for curious readers, we provide a theoretical analysis in the supplementary on addressable color gamut generated by conventional and multi-color holograms.
\vspace{\paraspace}

\paragraph{Image Quality.}
We compile \refTbl{image_quality} to provide an image quality comparison of multi-color scheme  against the conventional scheme.
In our assessments, we use commonly accepted image quality metrics of \PSNR, \SSIM, and \LPIPS~\cite{zhang2018unreasonable} (Readily available at \href{https://github.com/kaanaksit/odak}{\textbf{GitHub:odak}}~\cite{kavakli2022optimizing, kavakli2022introduction} and \href{https://github.com/photosynthesis-team/piq}{\textbf{GitHub:piq}} libraries~\cite{kastryulin2022piq}).
Our assessments compare the above two schemes for increasing intensity levels.
We invite readers also to observe the raw captures in our paper and supplementary.
\vspace{\paraspace}

\begin{table}[hbt!]
\caption{
Image quality evaluation of conventional and multi-color schemes for various levels of peak brightness.
Blue color indicates values for multi-color scheme.
}
\vspace{-2mm}
\resizebox{\columnwidth}{!}{%
\begin{tabular}{clccccc}
\toprule
\multicolumn{1}{l}{} & & \multicolumn{5}{c}{Peak Brightness} \\ \hline
\multicolumn{1}{l}{Scene} & Metrics   & $\times1.0$ & $\times1.5$ & $\times2.0$ & $\times2.5$ & $\times3.0$ \\ \hline
\multirow{3}{*}{\begin{tabular}[c]{@{}c@{}}AR \\ Glasses\\ (\refFig{brightness_increase})\end{tabular}} & PSNR (db) & \darkred{30.33}/\darkblue{29.92} & \darkred{23.82}/\darkblue{24.75} & \darkred{16.16}/\darkblue{22.39} & \darkred{12.16}/\darkblue{17.95}  & \darkred{9.66}/\darkblue{15.08} \\
 & SSIM      & \darkred{0.92}/\darkblue{0.91} & \darkred{0.86}/\darkblue{0.86} & \darkred{0.64}/\darkblue{0.82} & \darkred{0.40}/\darkblue{0.73}  &     \darkred{0.25}/\darkblue{0.65} \\
 & LPIPS     & \darkred{0.33}/\darkblue{0.33} & \darkred{0.38}/\darkblue{0.34} & \darkred{0.54}/\darkblue{0.36} & \darkred{0.64}/\darkblue{0.44}  &     \darkred{0.70}/\darkblue{0.50} \\ \midrule 

\multirow{3}{*}{\begin{tabular}[c]{@{}c@{}}Fruit \\ lady\\ (\refFig{beyond_18})\end{tabular}} & PSNR (dB) & \darkred{30.18}/\darkblue{29.65} & \darkred{22.19}/\darkblue{25.43} & \darkred{13.78}/\darkblue{22.32} & \darkred{9.40}/\darkblue{19.49}  & \darkred{6.80}/\darkblue{15.62} \\
 & SSIM      & \darkred{0.92}/\darkblue{0.90} & \darkred{0.81}/\darkblue{0.86} & \darkred{0.52}/\darkblue{0.81} & \darkred{0.30}/\darkblue{0.76}  &     \darkred{0.18}/\darkblue{0.67} \\
 & LPIPS     & \darkred{0.38}/\darkblue{0.36} & \darkred{0.47}/\darkblue{0.37} & \darkred{0.63}/\darkblue{0.41} & \darkred{0.70}/\darkblue{0.47}  &     \darkred{0.74}/\darkblue{0.55} \\ \midrule

\multirow{3}{*}{\begin{tabular}[c]{@{}c@{}}Dog\\ (\refFig{dynamic_scaling})\end{tabular}} & PSNR (dB) & \darkred{33.19}/\darkblue{31.03} & \darkred{23.87}/\darkblue{29.26} & \darkred{18.18}/\darkblue{26.68} & \darkred{15.23}/\darkblue{24.42}  & \darkred{13.17}/\darkblue{21.25} \\
 & SSIM      & \darkred{0.88}/\darkblue{0.79} & \darkred{0.81}/\darkblue{0.83} & \darkred{0.65}/\darkblue{0.80} & \darkred{0.50}/\darkblue{0.80}  &     \darkred{0.40}/\darkblue{0.76} \\
 & LPIPS     & \darkred{0.30}/\darkblue{0.33} & \darkred{0.37}/\darkblue{0.33} & \darkred{0.48}/\darkblue{0.36} & \darkred{0.54}/\darkblue{0.39}  &     \darkred{0.58}/\darkblue{0.43} \\

\bottomrule
\end{tabular}}
\label{tbl:image_quality}
\vspace{-5mm}
\end{table}

\paragraph{Ablation Study.}
We conduct an ablation study on our optimization model to identify the contribution of several components in our loss function and problem formulation.
Note that we conduct our study using actual results from our display hardware, but not simulations, as simulation models do not account for hardware imperfections, leading to perfect results in simulation but not in actual display hardware.
We provide the results from this study in \refTbl{ablation}, where we use the \PSNR, \SSIM, and \LPIPS image quality metrics.
In our ablation study, we remove one and only one component at each time.
There are four studies, and these studies involve removing double phase constrain (\refEq{db_phase}), total variation loss (\refEq{variation_loss}), laser loss (\refEq{laser_loss}), and running the complete optimization pipeline without removing any components.
We conduct this study by targeting $\times 1.8$ intensity values, using 1000 iteration steps and a 0.015 learning rate (Adam Optimizer~\cite{kingma2014adam}).
Our study suggests that TV loss and phase constrain are crucial in maintaining image quality. 
In addition, our practical observation suggests laser loss helps keep proper colors in reconstructed images.
\vspace{\paraspace}

\begin{table}[hbt!]
\caption{
Ablation Study for our multi-color holograms.
We remove only one component (not multiple) from our pipeline at each study and report image quality metrics. 
Without ``-'' component refers to the complete model.
}
\vspace{-2mm}
\centering
{%
\begin{tabular}{clccc}
\toprule
\multicolumn{1}{l}{Scene} & Without & PSNR (dB) & SSIM & LPIPS \\ \hline
\multirow{3}{*}{\begin{tabular}[c]{@{}c@{}}AR \\ Glasses \\ (\refFig{brightness_increase})\end{tabular}} & Phase Constrain & 11.48 & 0.32  & 0.72 \\
 & TV Loss & 13.72 & 0.57  & 0.55  \\
 & Laser Loss & 19.04 & 0.81  & 0.38 \\
 & - & 19.17 & 0.81  & 0.37 \\
\midrule

\multirow{3}{*}{\begin{tabular}[c]{@{}c@{}}Planets\\ (\refFig{extra_results})\end{tabular}} & Phase Constrain & 12.25 & 0.44  & 0.58 \\
 & TV Loss & 18.19 & 0.84  & 0.39  \\
 & Laser Loss & 23.82 & 0.81  & 0.42 \\
 & - & 26.27 & 0.64  & 0.42 \\

\midrule

\multirow{3}{*}{\begin{tabular}[c]{@{}c@{}}Candies\\ (\refFig{extra_results})\end{tabular}} & Phase Constrain & 8.41 & 0.13  & 0.98 \\
 & TV Loss & 12.39 & 0.44  & 0.74  \\
 & Laser Loss & 18.86 & 0.79  & 0.47 \\
 & - & 18.77 & 0.79  & 0.47 \\

\bottomrule
\end{tabular}
}
\label{tbl:ablation}
\vspace{-4mm}
\end{table}

\paragraph{Three-dimensional Images.}
The results we have shown for our method are two-dimensional.
However, our method can support three-dimensional scenes.
To enable three-dimensional support, $\lossTerm_{\text{image}}$ has to be replaced with a loss term supporting multiplanes (we use work the loss from work by Kavaklı et al.~\cite{kavakli2023realisticdefocus}).
In addition, the optimization formulated in \refEq{holohdr_optm} shall be applied to each plane, and the losses must be accumulated.
The results in \refFig{three_dimensional} and supplementary suggest that high-quality three-dimensional images are possible with our multi-color holograms.
\vspace{\paraspace}

\vspace{\paraspace}
\section{Discussion}
\label{sec:discussion}
Our multi-color holograms holds the potential to be an important tool for improving realism in the next-generation holographic displays.
However, there are various means to improve its performance, which we summarize in this section.
\vspace{\paraspace}

\begin{figure}[hbt!]
\centering
\includegraphics[width=0.95\linewidth]{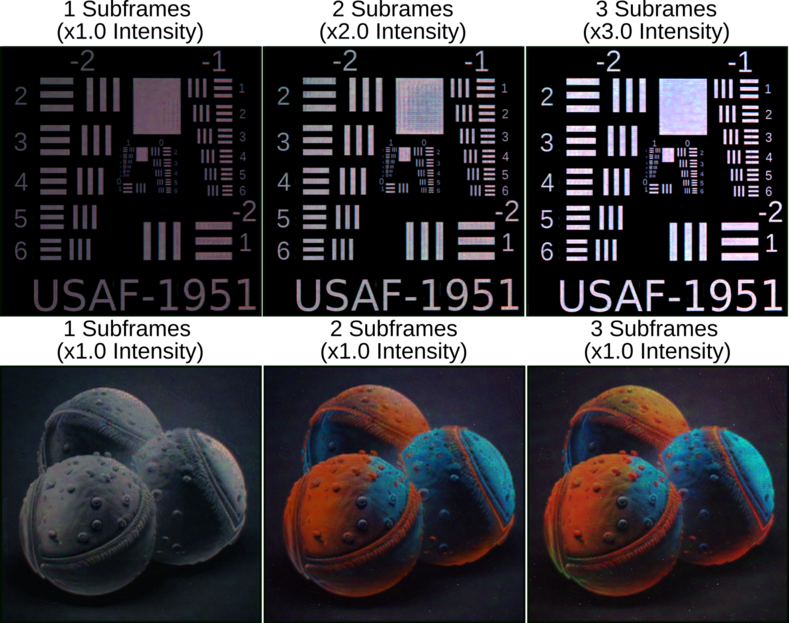}
\caption{
Using fewer subframes with our multi-color holograms.
The first row shows photographs of our multi-color hologram results with a peak brightness increase for a grayscale content (50 ms exposure).
The second row demonstrates the color reproduction quality increase for a full-color scene with the increasing number of subframes (Source image: Midjourney, Link: \href{https://github.com/complight/images}{\textbf{Github:complight/images}, 200 ms exposure}).
}
\label{fig:less_subframes}
\vspace{-7mm}
\end{figure}

\paragraph{Contrast and Dynamic Range}
Despite that our multi-color holograms achieve a peak brightness increase, it does not increase and could slightly decrease the contrast (\ie dynamic range). 
The reason is mainly two-fold.
Firstly, an \SLM's phase modulation is typically tuned to a specific wavelength.
Thus, when operating with three colors, \SLM performs with a full phase modulation range for one color while having limited phase modulations for the other two.
This loss of modulation accuracy leads to reduced diffraction efficiency and, consequently, lower contrast. 
Secondly, unlike conventional models focusing on achieving optimal response for a single color, each sub-frame in multi-color holograms needs to balance and ensure that the intensities for all three color channels approximate the desired scaled target image.
Thus, our method may choose intensities that could lead to slight deviations in color production.
We measure the Michelson contrast, $\frac{I_\mathrm{max} - I_\mathrm{min}}{I_\mathrm{max} + I_\mathrm{min}}$ to have a preliminary assessment of the situation. 
We measure for the highest and lowest brightness regions achieved in the top row example of \refFig{extra_results}.
We report the Michelson contrast as 0.94 for our method versus 0.99 for conventional in $\times 1.0$ brightness. 
But we also observe a trend with the increasing brightness scale.
For example, in $\times 1.8$ brightness case, the Michelson contrast is measured as 0.99 for multi-color and conventional cases. 
We speculate the loss of contrast issue may be mitigated by using loss functions dedicated to preserving contrast in the future.
\vspace{\paraspace}

\begin{figure}[hbt!]
\centering
\includegraphics[width=0.95\linewidth]{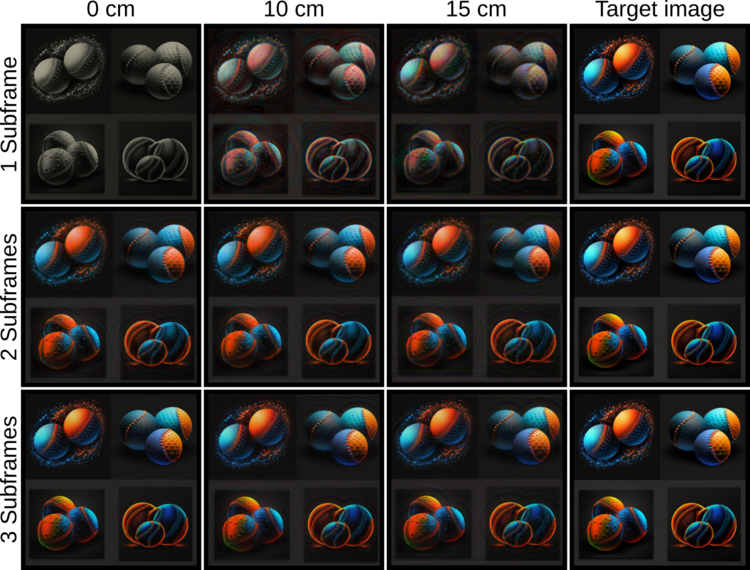}
\caption{
Given a target image (right) with $\times 1.0$ brightness and varying projection distances (from left to right), simulations of our multi-color holograms suggest an improvement in color reproduction capabilities when the projection distance prolongs, and may mean using fewer subframes to achieve the same image quality (Source image: Midjourney, Link: \href{https://github.com/complight/images}{\textbf{Github:complight/images}}).
}
\label{fig:propagation_distance}
\vspace{-6mm}
\end{figure}

\paragraph{Number of Subframes and color primaries}
In our evaluations, we use three subframes, $\numSubFrames = 3$.
However, as discussed in \refSec{method}, our method could also use a lower number of frames, $\numSubFrames \in \{1, 2\}$ (see \refFigFull{less_subframes}).
Fewer subframes can increase the refresh rate when monochrome and lower intensity target images are used (see \refFigFull{less_subframes} top row).
Similar to the work by Huang \etal~\shortcite{huang2017mixed}, using two subframes can also help display less colorful target images.
In addition, inspired by conventional displays with multiple color primaries~\cite{kauvar2015adaptive}, there could be a variant of our method with more color primaries or spatially structured illumination~\cite{huang2017mixed}, but holographic.
In this way, our multi-color hologram optimization could benefit from identifying the right set of color primaries or spatial distribution of the illumination source~\cite{jo2022multi}.
\vspace{\paraspace}

\paragraph{Long Propagation Distances.}
We report our results with images generated at the plane of \SLM for conventional and multi-color schemes.
When generating images away from an \SLM, the behavior of color reproduction can change noticeably due to the complex point spread functions induced at various propagation distances and wavelengths.
\refFigFull{propagation_distance} reveals such a case with simulated results generated at various distances from 0~cm to 15~cm for our hardware's color primaries.
An important observation from \refFig{propagation_distance} is longer propagation distances may help with accurate color reproduction using fewer subframes, as each pixel's color is now controlled by a larger subhologram, which endows more degree of design freedom.
This freedom stems from the varying size of diffracted light spread with the changing wavelength and distances.
At the extreme, $\numSubFrames = 1$, a long propagation distance of 15~cm could roughly match the color, promising the possibility of using our multi-color holograms to improve the frame rate.
However, the frame rate reduction process could also largely depend on targeted color content.
In practice, achieving good image quality without ringing artifacts at a long propagation distance remains a challenge for the state-of-the-art methods~\cite{kavakli2023realisticdefocus, shi2022end, choi2022time}.
In the future, expanding our multi-color holograms to support long propagation distances while exploring  alternative \SLM types~\cite{choi2022time} will be of great interest. 
Meanwhile, we find that optimizing a phase-only hologram without \DP constraint (direct phase coding) can produce visually similar results but with more noise (see supplementary). 
\vspace{\paraspace}

\paragraph{Hologram calculation speed.}
Convergence in our multi-color optimizations typically requires many steps (\eg 1000) and a small learning rate (\eg 0.015), leading to slow calculations (not interactive rate).
Specifically, a three plane multi-color hologram takes about ten minutes of optimization time with thousands steps on a NVIDIA RTX 3090.
However, a conventional hologram could calculate each subframe independently and concurrently with fewer steps (\eg 60) and memory demand.
Our multi-color optimizations could be formulated like a conventional hologram if required laser powers and targets for each subframe are known for a given content at the start of an optimization.
Our current multi-color optimizations could help generate a dataset where holograms with their corresponding laser powers and targets are provided.
Training a model with this dataset helps estimate the required laser powers at each subframe for a given target image before the optimizations.
\vspace{\paraspace}

\paragraph{Accounting Human Visual System.}
For spatial separation in color primaries in target scenes (e.g., a text where each color is represented with one color primary), our multi-color hologram solution will try to mimic conventional holograms (hologram per color primary).
Thus, the solution for such scenes could not benefit from brightness improvements while having artifacts degrading the image quality (see supplementary).
Our multi-color holograms assume that each color primarily contributes to only one perceived color.
As various combinations of color primaries can also isplay similar colors \cite{Schmidt2014NeurobiologicalHO}, accounting for \HVS in our method may help deliver perceptually accurate colors while relaxing the optimization, especially for targets with spatial color separation.
For further discussion on eyebox \cite{kim2022accommodative}, diffraction efficieny \cite{samanta2019study} and hardware-in-the-loop techniques \cite{peng2020neural, praneeth2020hardware, kavakli2022learned}, consult supplementary.

Holographic displays, has yet to be studied to support a similar feature.
For this purpose, we reimagine driving schemes for holographic displays.
Our solution offers a unique algorithmic change in calculating holograms.
This change also involves joint control of laser powers to illuminate the holograms more efficiently.
Our solution can help standard holographic displays to support higher intensity levels without using a more powerful laser.
\vspace{\paraspace}

\begin{acks}
We thank the knight of holography, Professor Byoungho Lee, for his services in the field~\cite{Park2023}.
The authors thank anonymous reviewers for their feedback.
Kaan Akşit is supported by the Royal Society's RGS/R2/212229 and Meta Reality Labs' inclusive rendering initiative.
Liang Shi is supported by Meta Research PhD Fellowship.
Hakan Urey is supported by the European Innovation Council’s HORIZON-EIC-2021-TRANSITION-CHALLENGES program (101057672) and Tübitak’s 2247-A National Lead Researchers Program (120C145).
\end{acks}
\vspace{\paraspace}

\begin{figure*}[hbt!]
\centering
\includegraphics[width=0.95\linewidth]{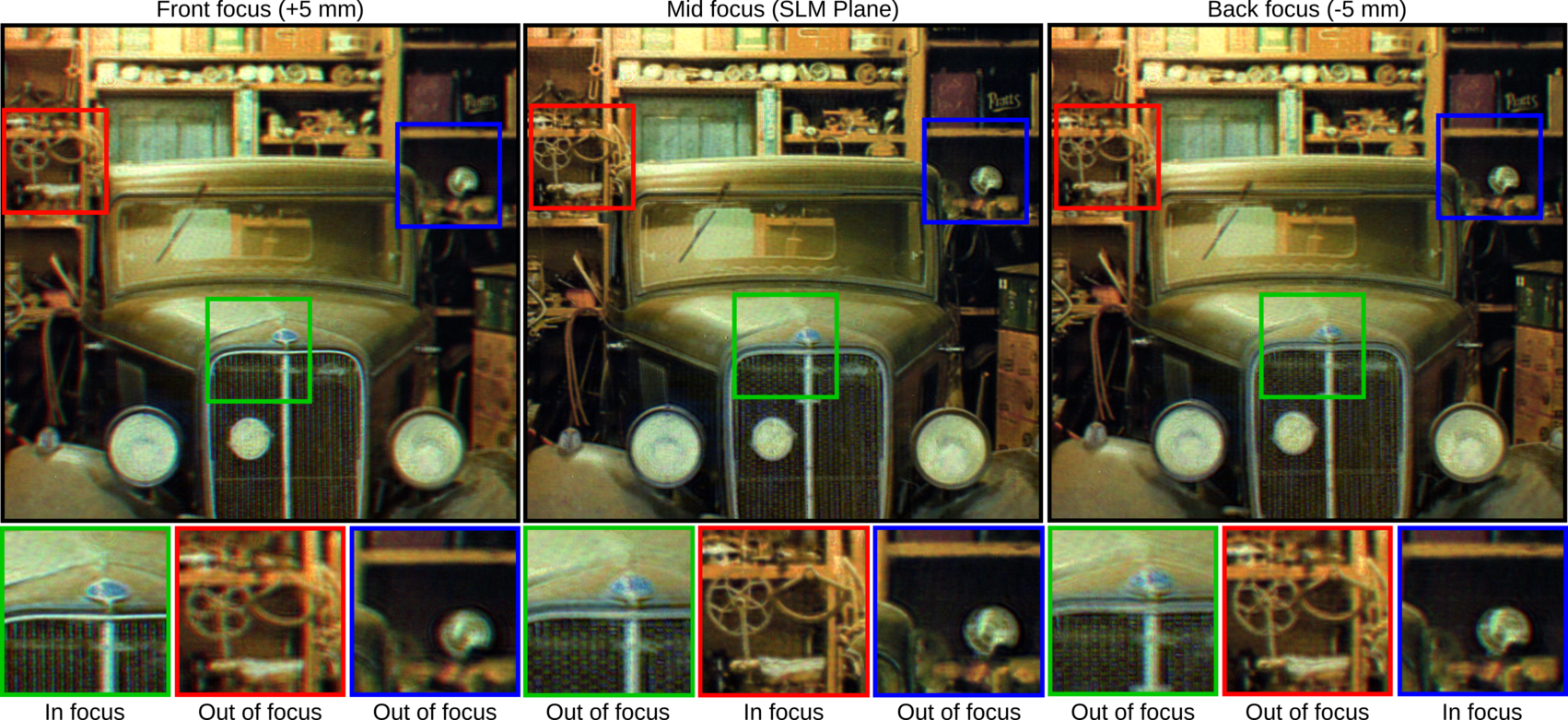}
\includegraphics[width=0.95\linewidth]{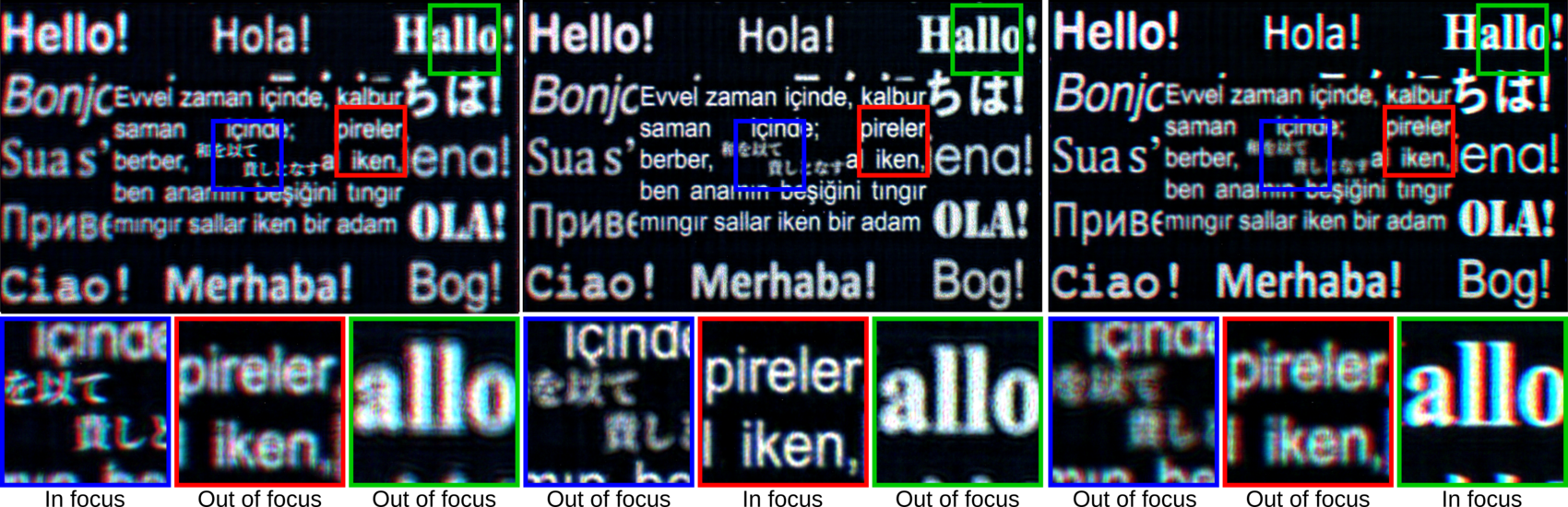}
\includegraphics[width=0.95\linewidth]{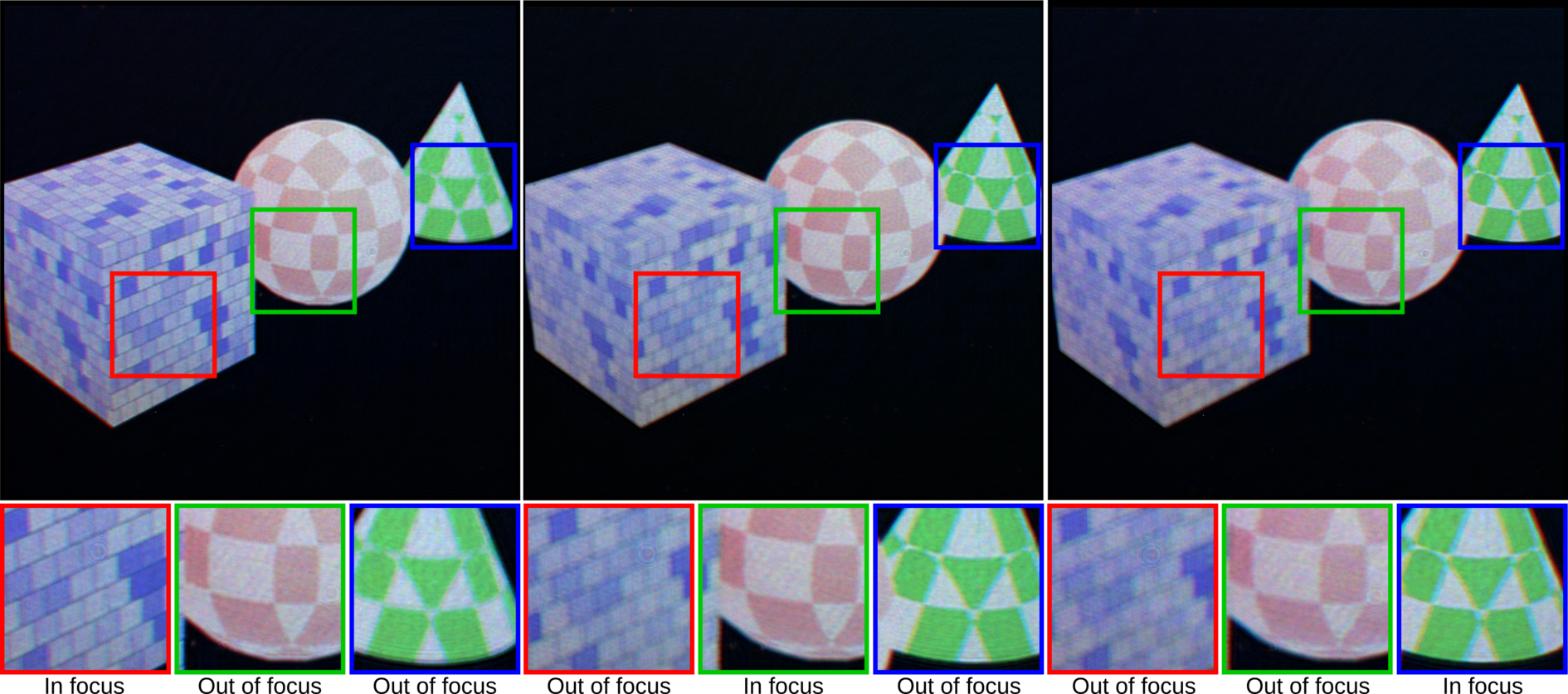}
\caption{
\3D scenes using our multi-color holograms. 
Each row shows a multiplane scene generated by our multi-color scheme with three focus planes. The targeted brightness level is $\times 1.8$ (Top image source: DIV2K~\cite{Agustsson_2017_CVPR_Workshops}, Other images (self-generated) source link: \href{https://github.com/complight/images}{\textbf{Github:complight/images}}, 150 ms exposure time).
}
\label{fig:three_dimensional}
\end{figure*}

\begin{figure*}[hbt!]
\centering
\includegraphics[width=1.0\linewidth]{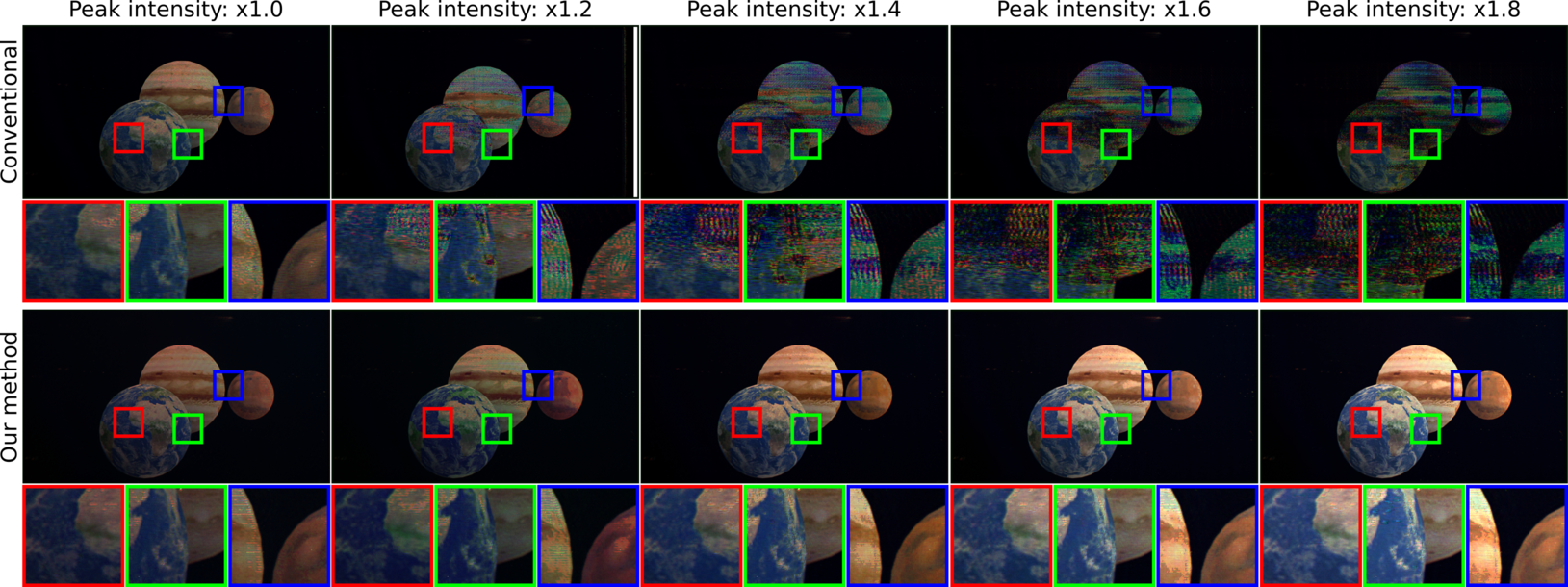}
\includegraphics[width=1.0\linewidth]{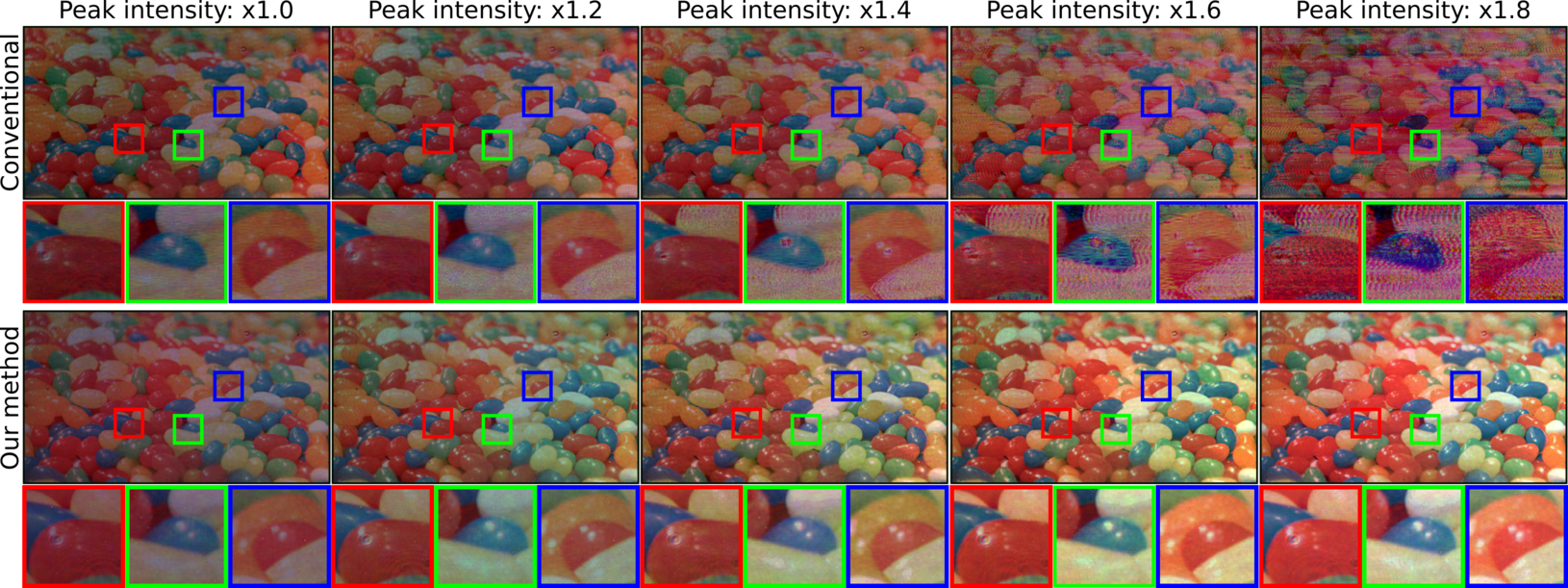}
\includegraphics[width=1.0\linewidth]{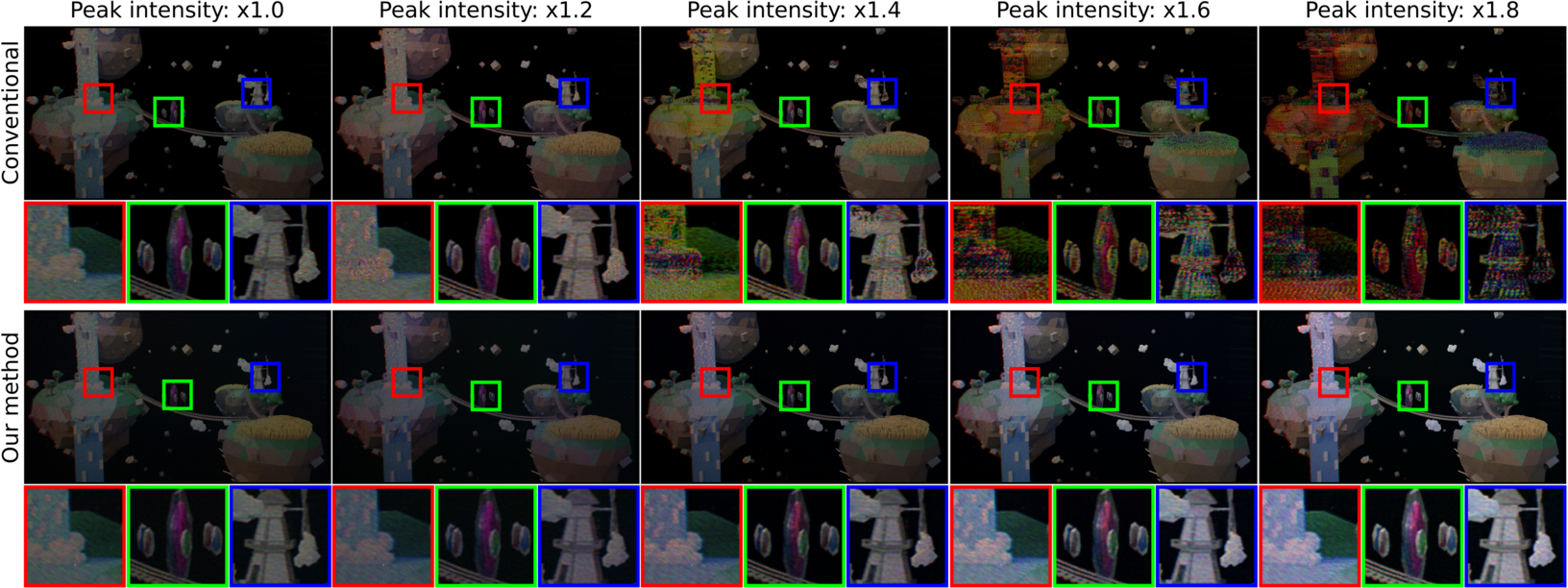}
\caption{
Increasing peak brightness levels with our multi-color holograms.
All photographs are captured at a 140 ms exposure.
Our multi-color holograms can enhance the peak brightness levels of the captures up to $\times1.8$ without artifacts or distortions, whereas conventional holograms fail to support beyond $\times1.0$ (Source link: \href{https://github.com/complight/images}{\textbf{Github:complight/images}}, Image Credits: Planets (self-generated), Candies \href{https://www.pexels.com/photo/a-close-up-shot-of-jelly-beans-5867973/}{\textbf{Pexels:Voyance Smith}}, Floating islands \href{https://www.cgtrader.com/free-3d-models/various/various-models/low-poly-floating-islands-3c579870-6f0a-41ff-90c4-f26e734a498a}{\textbf{CGTrader:ShalinSaju123}}, 140 ms exposure time).
}
\label{fig:extra_results}
\end{figure*}

\bibliographystyle{ACM-Reference-Format}
\bibliography{ref}

\AtEndDocument{

\section*{Supplementary material}

# Supplementary: Multi-color Holograms Improve Brightness in Holographic Displays

KORAY KAVAKLI, Koç University, Türkiye
LIANG SHI*, Massachusetts Institute of Technology, U.S.A.
HAKAN UREY, Koç University, Türkiye
WOJCIECH MATUSIK, Massachusetts Institute of Technology, U.S.A.
KAAN AKŞIT*, University College London, United Kingdom

## 1 HOLOGRAPHIC DISPLAY HARDWARE

Our holographic display helps us to assess the image quality in our multi-color holograms and is also helpful for comparisons against conventional holograms. Figure 1 provides a photograph of our holographic display prototype.

Here, we provide a list of components used in our holographic display. It starts with a fiber-coupled multi-wavelength laser light source, LASOS MCS4, which combines three laser light sources peeking at 473 nm, 515 nm, and 639 nm. Two ESP32 boards control our multi-wavelength laser light source LASOS MCS-4. For accurate power control, we relied on Digital-to-Analog Converters (DAC) that are available on ESP32 boards. There is a pinhole aperture, Thorlabs SM1D12, after some distance concerning the fiber tip. This aperture helps us limit the diverging beams from our fiber. After this pinhole aperture, there is a linear polarizer, Thorlabs LPVISE100-A, which enables a polarization state aligned with our phase-only Spatial Light Modulator's fast axis (SLM) for light beams. Linearly polarized light beams reach our phase-only SLM, Holoeye Pluto-VIS, and get modulated with the optimized phase pattern. The phase-modulated beam arrives at a 4f imaging system composed of two 50 mm focal length achromatic doublet lenses, Thorlabs AC254-050-A, and a pinhole aperture, Thorlabs SM1D12, removing unmodulated and undiffracted light. In our experiments, we used a Ximea MC245CG-SY camera to capture the image reconstructions. We place our camera on an X-stage (Thorlabs PT1/M travel range: 0-25 mm, precision: 0.01 mm) and move it back and forth to capture photographs from various depth levels.

*Discussion on the Off-Axis Holographic Display Prototype.* The hardware prototype we employed for our experiments exhibits an off-axis configuration, wherein the direction of the diffracted beam is wavelength-dependent, resulting in chromatic aberrations as images move away from the Spatial Light Modulator (SLM) plane. These effects are evident in our 3D results, as depicted in Figure 9 of our manuscript. Specifically, both the front and back planes positioned 5 mm away from the SLM plane are susceptible to this issue. In contrast, on-axis configured holographic display prototypes hold the potential to overcome these aberrations by forming images on the zeroth diffraction order. However, it is crucial to note that on-axis systems inherently suffer from low contrast and limited dynamic range due to the presence of unmodulated beams. For off-axis holographic systems, there exist several avenues to mitigate the impact of chromatic aberrations:

---

*denotes corresponding authors

---

Authors' addresses: Koray Kavaklı, Koç University, Türkiye, kavakli@ku.edu.tr; Liang Shi, Massachusetts Institute of Technology, U.S.A., liang@mit.edu; Hakan Urey, Koç University, Türkiye, hurey@ku.edu.tr; Wojciech Matusik, Massachusetts Institute of Technology, U.S.A., wojciech@mit.edu; Kaan Akşit, University College London, United Kingdom, k.aksit@ucl.ac.uk.

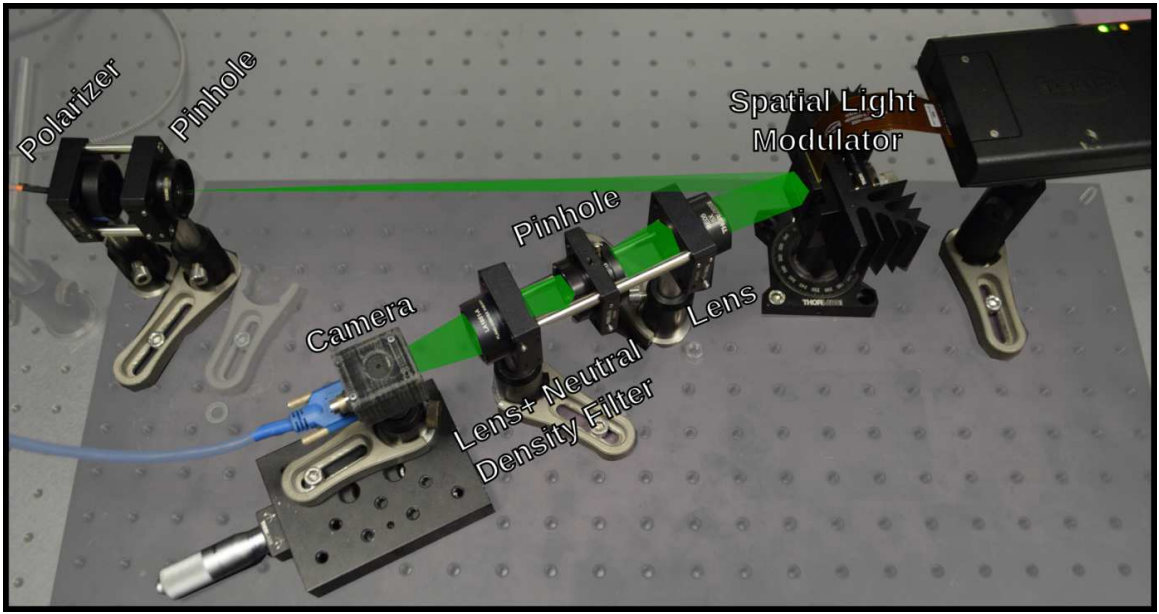

Fig. 1. A photograph showing our holographic display prototype used in assessing our multi-color holograms.

*Blazed Gratings.* In theory, using blazed grating terms instead of linear phase gratings presents a viable solution to address the chromatic aberrations in off-axis systems. By incorporating blazed gratings, the misalignment of diffracted beams can be mitigated. However, a significant challenge lies in the practical implementation, as it necessitates sampling at a subpixel level of the SLM for each wavelength. Consequently, directly applying wavelength-dependent blazed gratings to rectify the misalignment of diffracted beams in holographic displays proves to be complex.

*Improved Illumination Optics.* An alternative approach involves precision optical design of the illumination optics. One potential method to rectify the misalignment of diffracted beams in off-axis holographic displays entails utilizing a spatially decentralized laser source with varying wavelengths, slightly tilted concerning their half-order diffraction angle. This solution requires three single-mode fibers, each carrying a distinct wavelength source. An alternative to avoid using three separate beams coupled to individual laser diodes involves employing specially designed optical hardware with prisms and RGB color filters. This setup efficiently filters and refracts the illuminated white light source into three primary colors, each following different paths [12] at the expense of light efficiency. Both of these methods hold promise in reducing chromatic aberrations and facilitating the overlapping of modulated hologram beams in the Fourier aperture plane.

## 2 EXTENDED LASER LOSS

In the main body of our manuscript, the laser loss of our method is described as

$$L_{\text{laser}} = \sum_{p=1}^{3} \left( \left( \sum_{t=1}^{T} l_{(p,t)}^2 \right) - \max(I_p)s \right)^2. \tag{1}$$

This laser loss described in Eq. (1) could be further extended as $L'_{\text{laser}}$ to avoid laser powers converging to zero or distributing unevenly. Our extension to Eq. (1) is as follows,

$$
\begin{aligned}
L'_{\text{laser}} = L_{\text{laser}} &+ \cos\left(\min\left(\sum_{p=1}^{3} l^2_{(p,t)}\right)\right) \\
&+ \cos\left(\min\left(\sum_{t=1}^{T} l^2_{(p,t)}\right)\right) \\
&+ \left((Ts) - \sum_{p=1}^{3}\sum_{t=1}^{T} l^2_{(p,t)}\right)^2,
\end{aligned}
\tag{2}
$$

In this extension, the first component we have to observe is as follows:

$$
\cos\left(\min\left(\sum_{p=1}^{3} l^2_{(p,t)}\right)\right)
\tag{3}
$$

which helps to regularize the minimum value in the sum of laser powers across color primaries. This way, we ensure that the minimum total power for each color primary is non-zero. The second component,

$$
\cos\left(\min\left(\sum_{t=1}^{T} l^2_{(p,t)}\right)\right),
\tag{4}
$$

help ensure that the total power for each frame is non-zero. These two components avoid hitting zero in terms of power for all frames and colors. The third and last component,

$$
\left((Ts) - \sum_{p=1}^{3}\sum_{t=1}^{T} l^2_{(p,t)}\right)^2,
\tag{5}
$$

encourages the optimizer to meet the sum of laser power as large as the peak brightness level times the number of subframes. Thus, encouraging the optimized laser powers to meet the brightness demand for a given target.

## 3 OPTICAL BEAM PROPAGATION

Light transport models play a critical role in simulating coherent light used in holographic display applications (and more). We typically represent phase-only holograms used in a holographic display using a two-dimensional array filled with phase values ranging from $-\pi$ to $\pi$. We can also describe such a phase-only hologram in a complex notation, $O_h = e^{j\phi(x,y)}$, where $\phi$ represents the phase delay introduced by each pixel at a phase-only hologram. Holographic displays typically represent holograms, $O_h$, with programmable SLMs. A coherent beam $U_i$, again represented as a two-dimensional array, illuminates the phase-only hologram, $O_h$. Note that $U_i$ is an oscillating electric field described as $U_i = A_0 e^{j(k\vec{r} + \phi_0(x,y))}$, where $A_0$ represents the amplitude of the optical beam, $k$, means the wavenumber that can be calculated as $\frac{2\pi}{\lambda}$, $\lambda$ represents the wavelength of light, and $\phi_0$ represents the initial phase of the optical beam. $A_0$ is often considered as $A_0 = 1$ for an ideal collimated beam, while $\phi_0$ is assumed to be a two-dimensional array filled with random values between zero to $2\pi$. Finally, leading to simplification of $U_i$ as $e^{j\phi_0}$. In simple terms, as $U_i$ illuminates $O_h$, $U_i$ by modulated with $O_h$, forming a new modulated beam $U_m$ that is calculated as

$$
U_m = U_i O_h = e^{j(\phi(x,y) + \phi_0(x,y))}.
\tag{6}
$$

We form the reconstructed images at various depths as the modulated beam, $U_m$, propagates in free space away from the hologram plane (SLM plane). This propagation of optical beams from one plane to another follows the theory and method introduced by Rayleigh-Sommerfeld diffraction integrals [4]. This diffraction integral's first solution, the Huygens-Fresnel principle, is expressed as follows:

$$u(x, y) = \frac{1}{j\lambda} \int \int u_0(x, y) \frac{e^{jkr}}{r} \cos(\theta) dx dy, \tag{7}$$

Where the resultant field, $U(x, y)$, is calculated by integrating over every point across the hologram plane, $U_0(x, y)$ represents the optical field in the hologram plane for every point across XY plane (perpendicular to propagation direction), $r$ represents the optical path between a selected point in hologram plane and a selected point in target plane, $\theta$ represents the angle between these points. The angular spectrum method, an approximation of the Huygens-Fresnel principle, is often simplified into a single convolution with a fixed spatially invariant complex kernel, $h(x, y)$ [17],

$$u(x, y) = u_0(x, y) * h(x, y) = \mathcal{F}^{-1}(\mathcal{F}(u_0(x, y))\mathcal{F}(h(x, y))). \tag{8}$$

In our implementations, we use a differentiable implementation of the light transport model found in Eq. (8), which we import from **GitHub:odak** [6, 9].

## 4 GRADIENT DESCENT OPTIMIZATION WITH DOUBLE PHASE CONSTRAINT

Our method aims to generate images that remain at the proximity of an SLM following the literature [13, 16]. Images in the proximity of an SLM are known for their high image quality. In this region, the light propagation distances $r$ are typically a few millimeters. This region's most common phase-only hologram encoding method is the Double Phase (DP) [5] approach. DP method decomposes a complex field into a phase-only hologram. When optimizing holograms with Gradient Descent (GD) optimization, it is possible to introduce DP encoding into the optimization pipeline by defining a phase constraint [7]. Optimization can provide DP-encoded optimized holograms by constraining the phase updates of GD, $\phi$ of $O_h$. This constraint for $\phi$ can be written as a decomposition:

$$\phi_0 = \phi - \bar{\phi}$$
$$\phi_{\text{low}} = \phi_0 - \text{offset}$$
$$\phi_{\text{high}} = \phi_0 + \text{offset}$$
$$x_{\text{even}}, y_{\text{even}} \in \{0, 2, 4, 6, \dots\}$$
$$x_{\text{odd}}, y_{\text{odd}} \in \{1, 3, 5, 7, \dots\}$$
$$\phi[x_{\text{even}}, y_{\text{even}}] = \phi_{\text{low}}[x_{\text{even}}, y_{\text{even}}]$$
$$\phi[x_{\text{odd}}, y_{\text{odd}}] = \phi_{\text{low}}[x_{\text{odd}}, y_{\text{odd}}]$$
$$\phi[x_{\text{even}}, y_{\text{odd}}] = \phi_{\text{high}}[x_{\text{even}}, y_{\text{odd}}]$$
$$\phi[x_{\text{odd}}, y_{\text{even}}] = \phi_{\text{high}}[x_{\text{odd}}, y_{\text{even}}]$$
$$O_h \leftarrow \phi, \tag{9}$$

, where offset is a variable to be optimized.

## 5 COMPARISON BETWEEN CONVENTIONAL AND MULTI-COLOR HOLOGRAM SCHEMES

This section compares additional results from conventional and multi-color schemes while targeting up to ×1.8 peak intensity levels. Actual photographs of these comparisons are readily available in Fig. 2 and Fig. 3. In Fig. 4, we conducted a comparative analysis between multi-color holograms

and conventional holograms with an higher laser power outputs. The evaluation of image quality indicates a significant drop in the case of conventional holograms with higher laser power output as the background noise increases in the images. Multi-color holograms also achieves similar or higher image quality with respect to conventional holograms at laser power outputs of ×1.5 and ×2.0, all while upholding the desired peak intensity levels. We also provide a pseudo-code for our conventional optimization routine, as in Listings 1. Readers can find our double-phase implementation structure in Sec. 4.

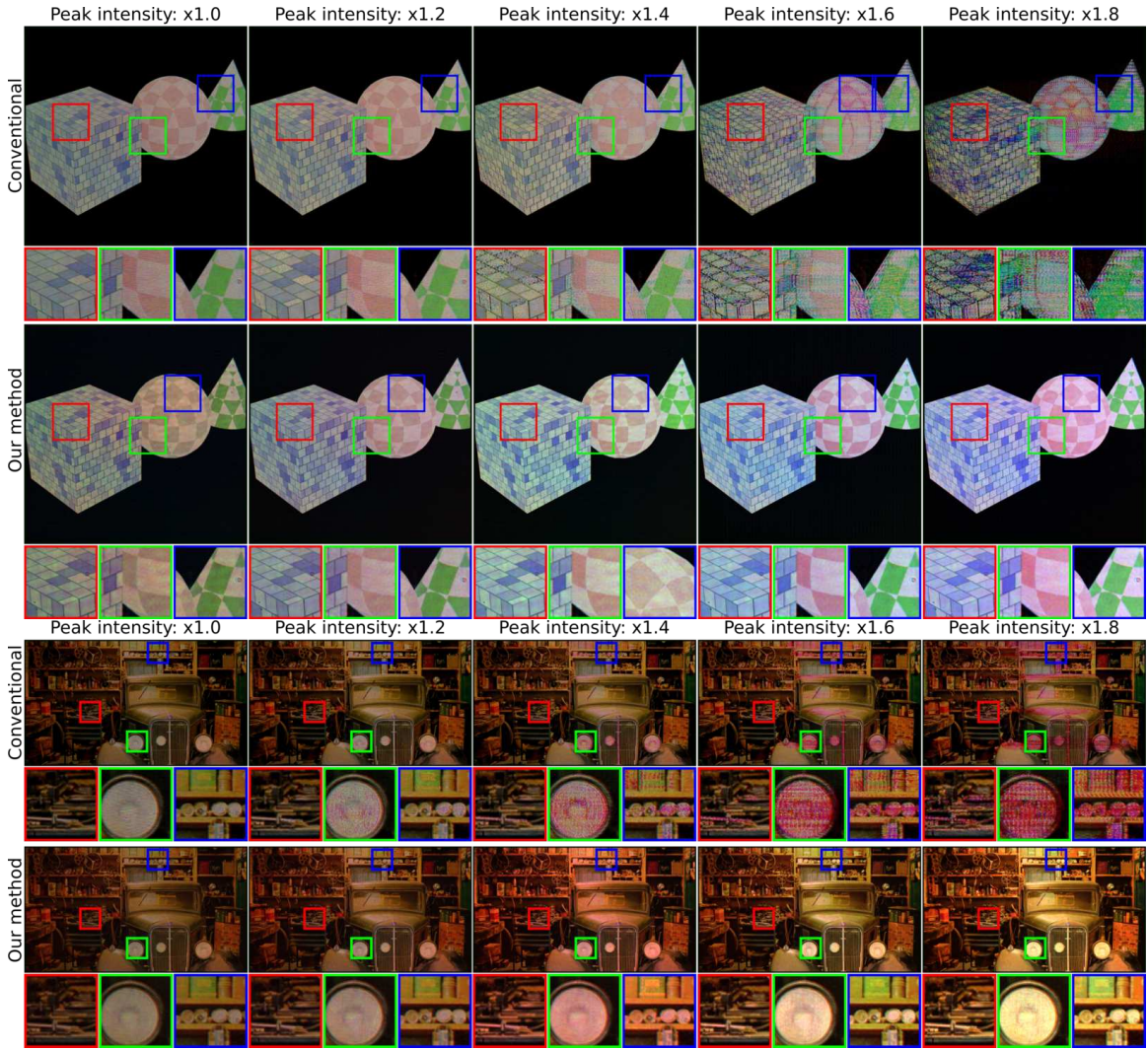

Fig. 2. Additional comparison between conventional and multi-color schemes. Photographs show that multi-color holograms can enhance the peak brightness levels of the captures up to ×1.8 without artifacts or distortions. All the images are generated on the Spatial Light Modulator (SLM) plane and captured with 140 ms exposure time using our holographic display. The conventional scheme fails to generate holograms that can target beyond ×1.0 brightness levels. (140 ms exposure).

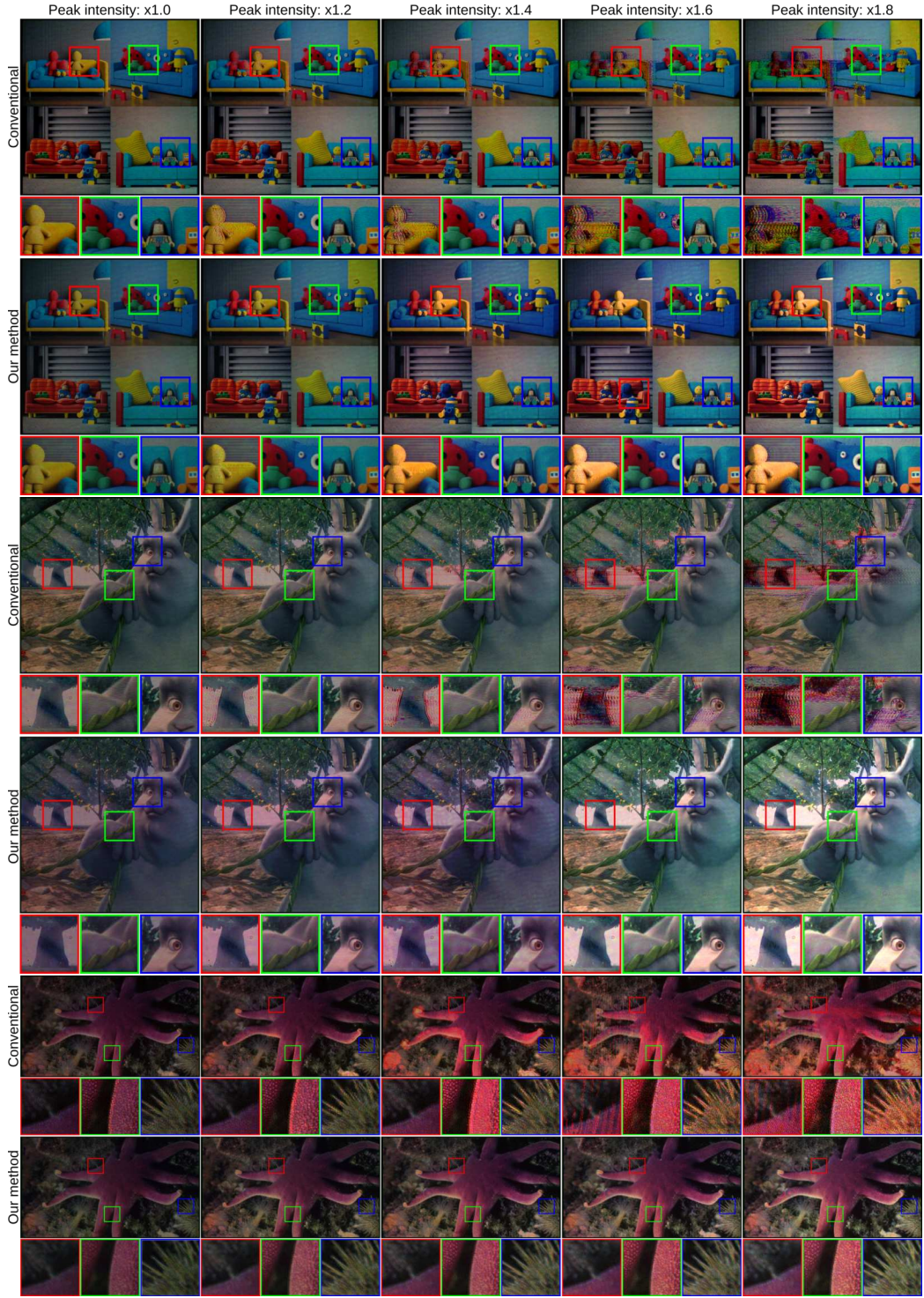

Fig. 3. Increasing peak brightness levels with our multi-color holograms. Photographs show that our method can enhance the peak brightness levels of the captures up to ×1.8 without artifacts or distortions. In contrast, the conventional hologram fails to support beyond ×1.0 (140 ms exposure).

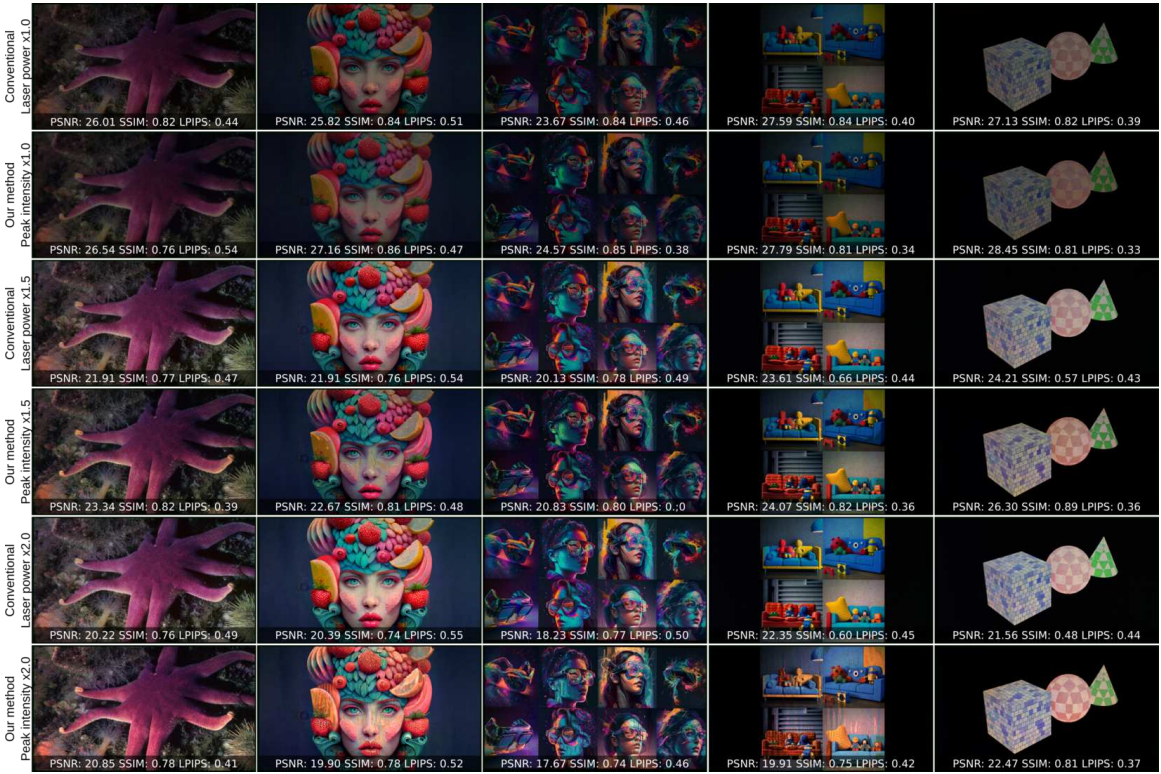

Fig. 4. Impact of increased laser power on conventional holograms. Photographs in the figure demonstrate the effect of elevated laser power on conventional holograms. Image quality assessment reveals a significant reduction in quality due to amplified background noise for conventional. Whereas, multi-color holograms achieve comparable image quality to conventional holograms at laser power outputs of ×1.5 and ×2.0 , while maintaining desired peak intensity levels. Notably, multi-color holograms exhibit better or same image quality metrics when targeting 1.5x intensity, while showing slight color balance deviations when targeting 2.0x peak intensities. (140 ms exposure time).

## 6 SIMULATED RESULTS

In Fig. 5, we provide full color simulated results along with the each individual reconstruction for each frame using our multi-color holograms when the target peak brightness is ×1.8.

## 7 BEYOND ×1.8 PEAK BRIGHTNESS LEVELS

Our method can also target peak brightness levels beyond ×1.8 at the expense of color integrity. Beyond this threshold, our captures from our multi-color holograms resemble image artifacts similar to conventional holograms beyond ×1.0. In Fig. 6, we provide additional captures acquired from our holographic display when the target peak intensity levels exceed ×1.8 for multi-color hologram scheme.

## 8 MULTIPLANE MULTI-COLOR HOLOGRAMS

We provide additional three dimensional scene captures of the multiplane images that are generated with our multi-color hologram scheme in Fig. 7 and Fig. 8.

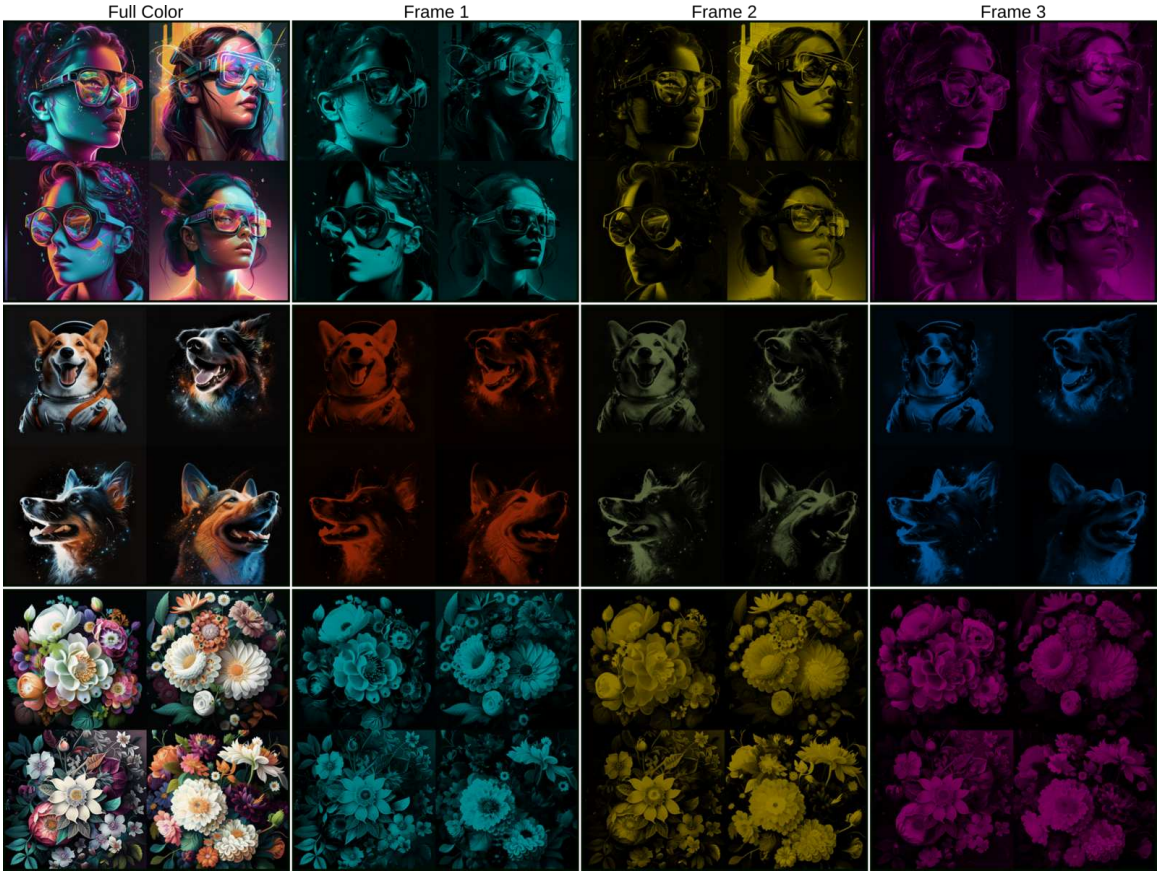

Fig. 5. Simulated results showing full color and each time frame reconstructions that are generated with our multi-color holograms when targetting ×1.8.

## 9   DYNAMIC INSENSITY SCALING

Our method's dynamic peak intensity scaling option allows increasing the peak brightness of the reconstructed images by simply setting a set of desired image qualities (image loss thresholds). We optimize the example experimental captures from our holographic display in Fig. 9 using a multi-color hologram dynamic intensity scale for various image qualities. From our experimental assessments in Fig. 9 and simulation-based assessments in Fig. 10, we conclude that aiming for lower image quality (higher image loss thresholds) enables a higher peak brightness at the expense of visual artifacts and color variations.

## 10   DIRECT PHASE REPRESENTATION

Our multi-color hologram scheme also supports direct phase encoding. We provide early image quality assessments showing slighly noisy results by switching to direct encoding from DP encoding. Direct phase results are presented in Fig. 11. We share these results to provide guidance for this important debate in the community regarding DP vs direct encoding.

## 11   SPATIAL COLOR SEPARATION AND CONTENT DEPENDENCY

We discuss at the manuscript of our paper that multi-color holograms could not benefit from brightness improvements and could have visual distortions when the target scene contains spatial

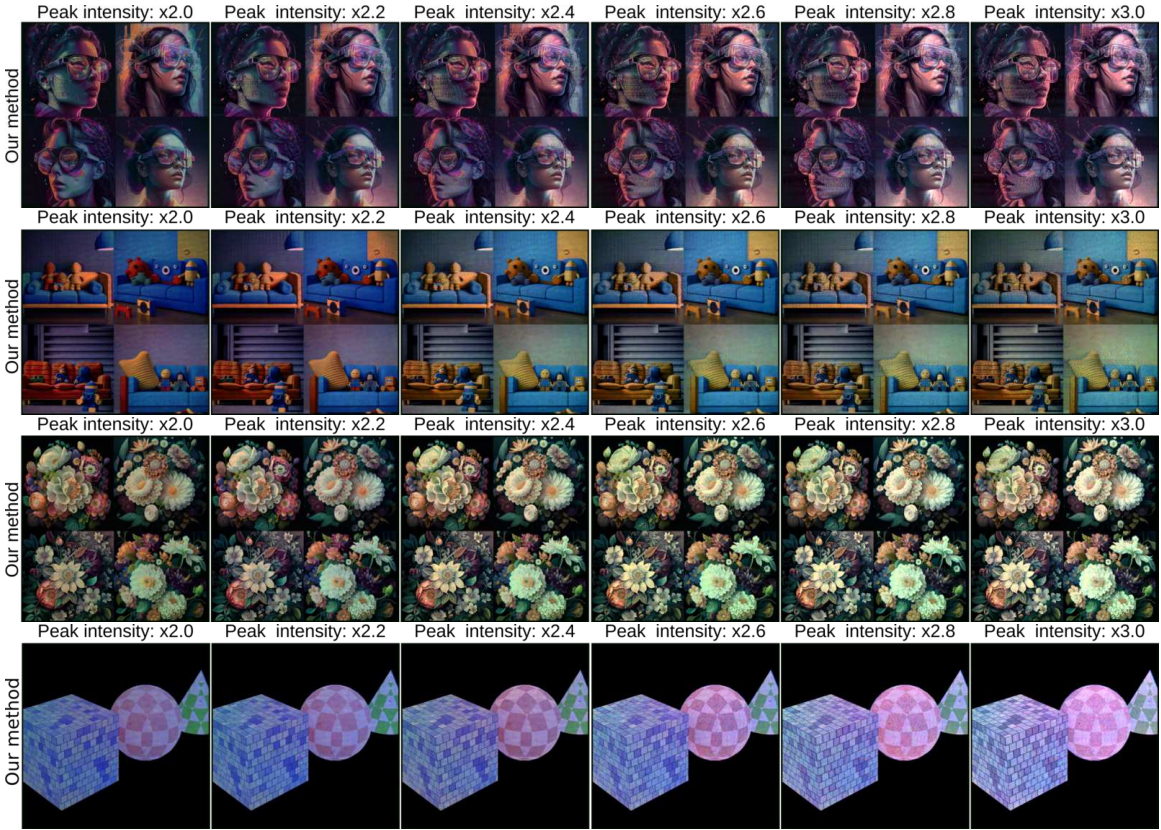

Fig. 6. Targeting beyond ×1.8 peak brightness levels. Photographs showing our multi-color holograms generating higher brightness beyond ×1.8 (50 ms exposure time)

color separation (e.g. each letter of a text is dedicated to one color primary). Although this is a very unlikely case in a display application, we provide a sample of the issue in Fig. 12.

## 12   REDUCING THE COST OF ILLUMINATION SOURCE

Our method can enable the use of low-power lasers in holographic displays and could provide a benefit on cost reduction for the development of holographic displays in the future. To illustrate this, let's consider two different laser diodes that are commercially available in Thorlabs: the HL6322G and the HL6312G. The HL6322G is a 15mW laser diode that costs $77.45, while the HL6312G is a 5mW laser diode that costs $24.45. The HL6322G is three times more expensive than the HL6312G.

## 13   COLOR PERFORMANCE

Our method increases the brightness of holographic displays by utilizing its light sources more effectively through multi-color holograms. We find that the effectiveness of this brightness enhancement is content dependent. In other words, the degree of brightness enhancement could change depending on the target content. In some cases, going beyond what brightness level content allows could cause visual artifacts and a mismatch in color. To understand how, truthfully, our method could, in principle, maintain colors with varying brightness levels, we conduct a series of experiments in simulations. We chose a set of target images and optimized their multi-color holograms for varying brightness levels, ×1.0, ×1.2, ×1.4, ×1.6, and ×1.8. We compare simulated reconstructions of these optimized multi-color holograms with their corresponding targets at three

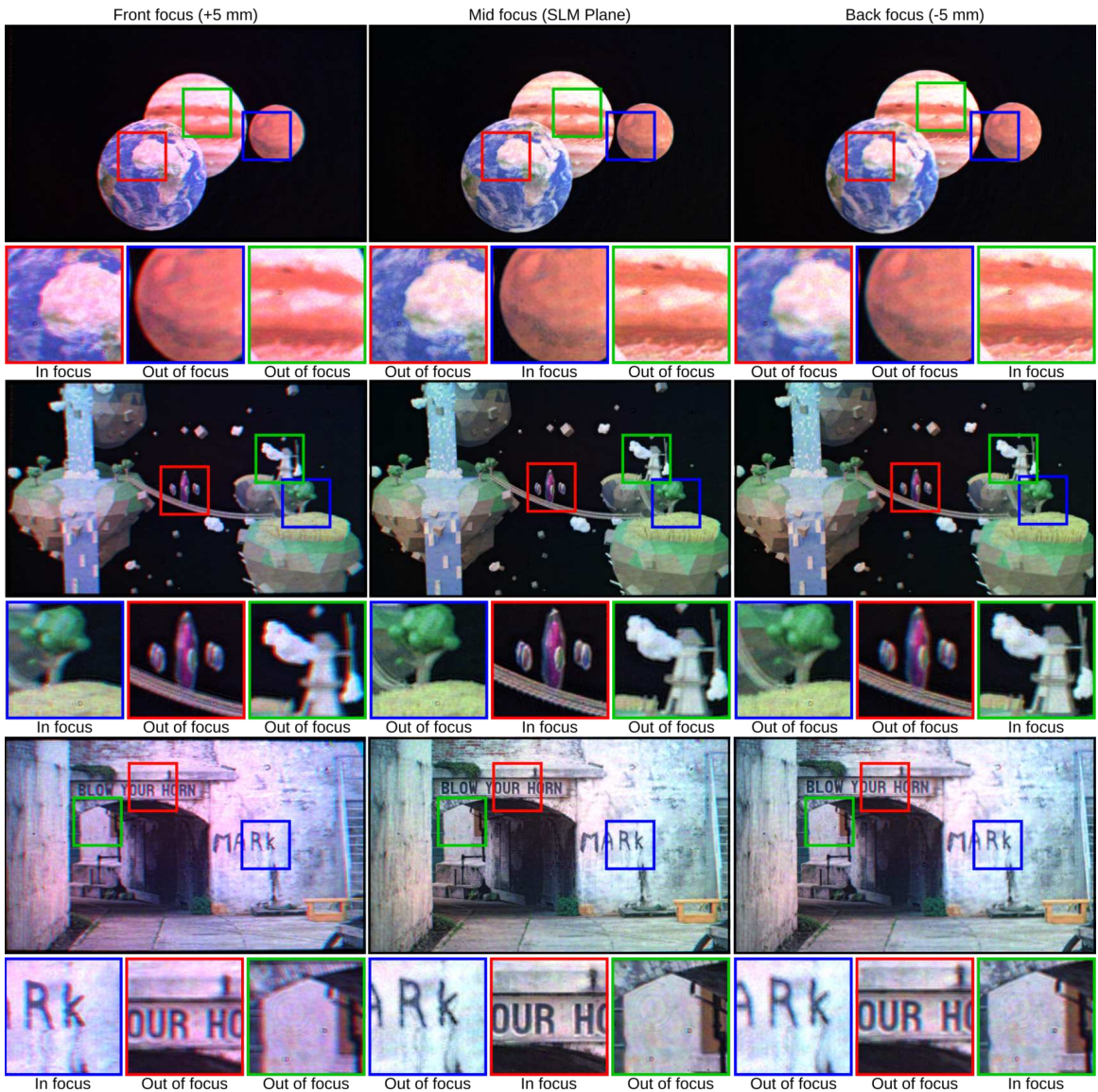

Fig. 7. Three dimensional scenes using our multi-color holograms. Photographs show a multiplane image generated by our multi-color hologram scheme with three focus planes. The targeted brightness level is ×1.8. (150 ms exposure time)

separate depth levels, near, mid, and far, corresponding to 0 *mm*, 5 *mm*, and 10 *mm* away from SLM. Given the three color primaries used in our holographic display prototype, 473 *nm*, 515 *nm*, and 639 *nm*, we convert these reconstructions and their targets to trichromat sensations in the LMS cones of Human Visual System (HVS). We follow the exact conversion from RGB to LMS highlighted in the work by [3]. We calculate a Euclidean and Chamfer distance for each pair in LMS space to understand how close the simulated reconstructions are to their targets. The results of these comparisons in LMS space are provided in Fig. 13, Fig. 14, Fig. 15, and Fig. 16. We observe from these assessments that our method mostly maintains color consistency with target scenes. However, higher brightness levels in some content could lead to perceptible color differences between target scenes and their corresponding solutions generated by our method (e.g. Fig. 16).

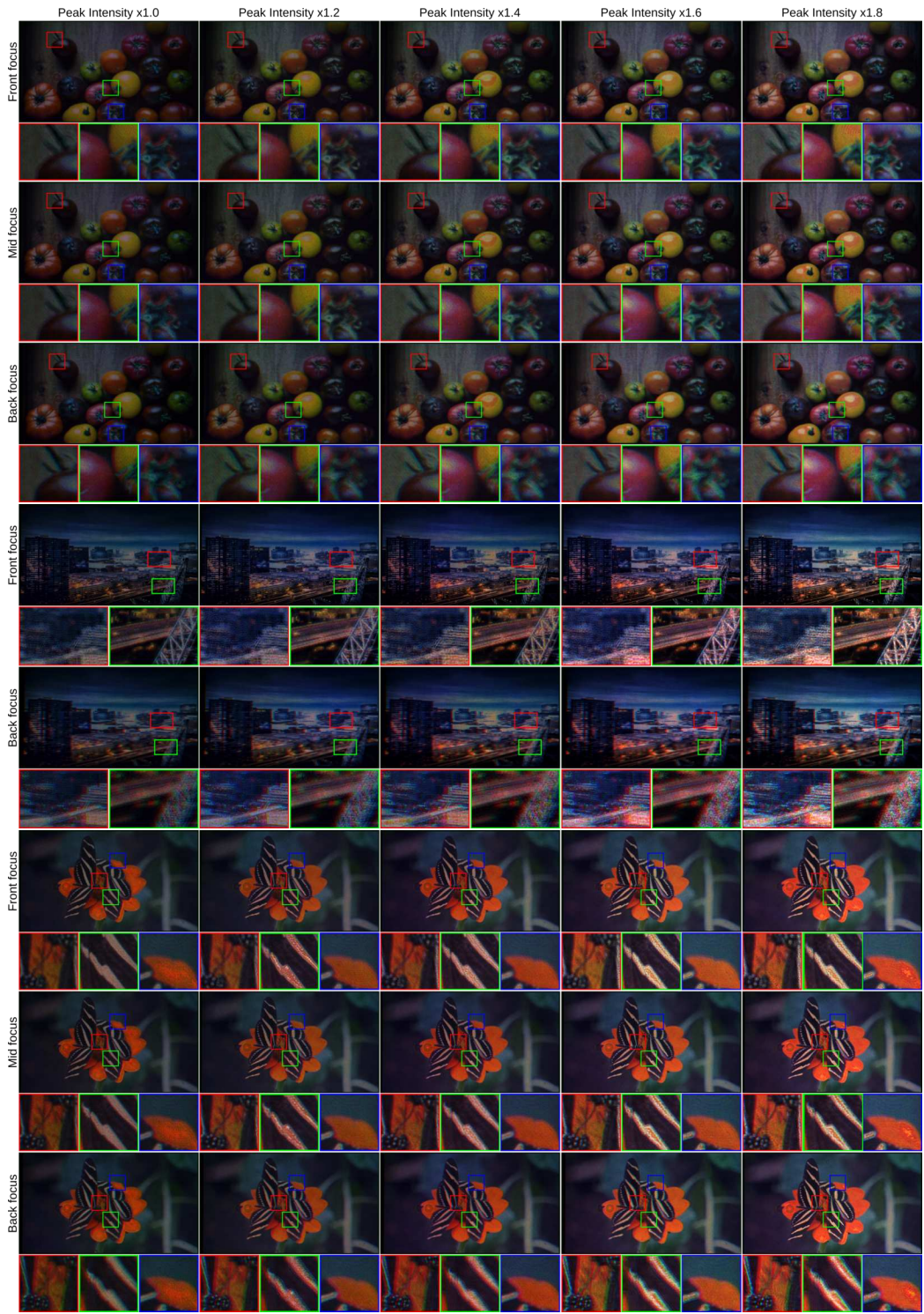

Fig. 8. Three dimensional scenes using our multi-color holograms. Photographs show a multiplane image generated by our multi-color hologram scheme with three focus planes. The targeted brightness level is ×1.8. (150 ms exposure time)

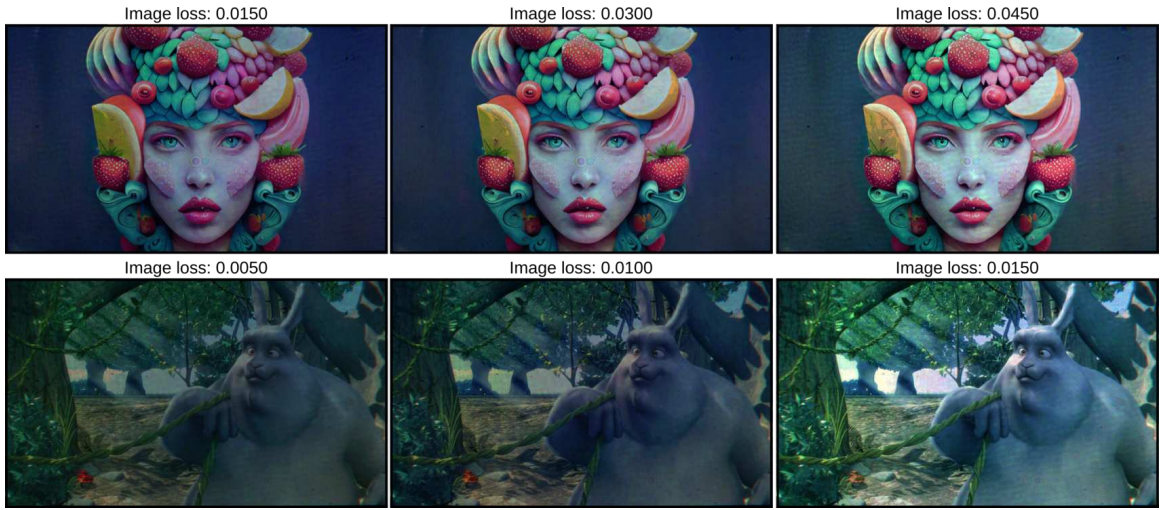

Fig. 9. Dynamic intensity scaling using our method. Photographs of the images that are generated with different image loss threshold by our multi-color hologram dynamic intensity scaling. The optimized peak intensity levels for the first row: ×1.61, ×1.87 and ×2.04 and the second row: ×1.61, ×1.27 and ×1.45 is ×2.23. (100 ms exposure time)

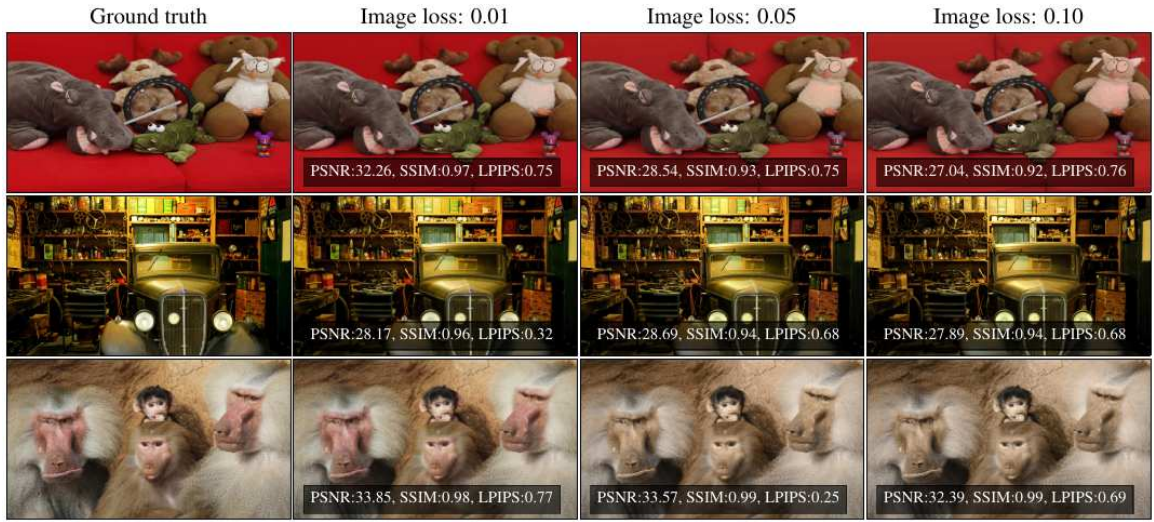

Fig. 10. Dynamic intensity scaling using our method. Simulated images that are generated with different image loss threshold by our multi-color hologram dynamic intensity scaling. The optimized peak intensity levels for the first row: ×1.46, ×2.17 and ×2.57, the second row: ×1.76, ×2.24 and ×2.47, and the third row: ×2.1, ×3.13 and ×3.36.

## 14 ITERATION COUNT

The number of iterations used in our method's optimization determines the final outcome's image quality and color performance. We compiled a figure as in Fig. 17 to demonstrate this relation. In this figure, we target Three-Dimensional (3D) scenes and rely on the exact configuration in our optimization code (e.g., defocus blur size in target images, propagation distances), where we choose a 1 cm volume 0.5 cm away from an SLM. Thus, we add a series of images in Fig. 17 to

Peak Intensity x1.0          Peak Intensity x1.5          Peak Intensity x2.0

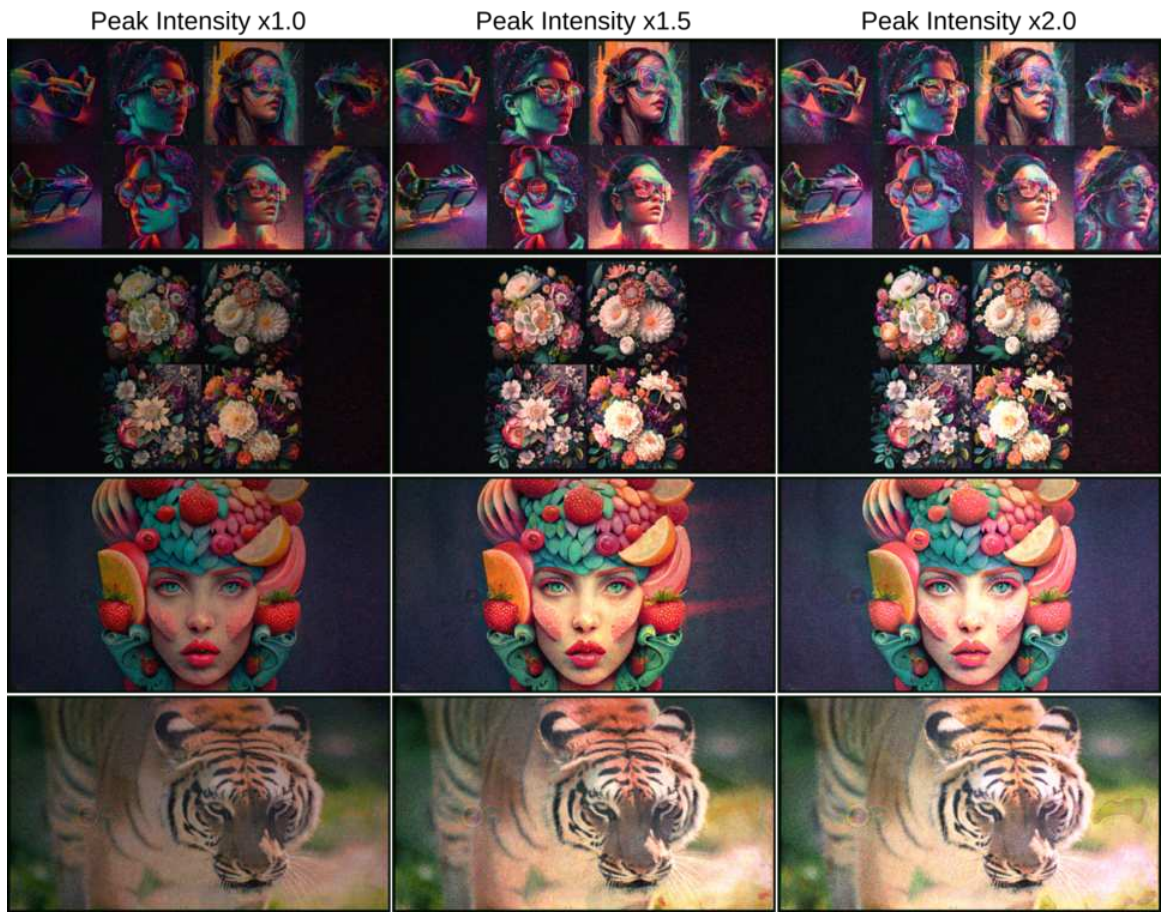

Fig. 11. Captured images for direct phase holograms at Spatial Light Modulator (SLM) plane. Photographs showing our multi-color holograms with direct phase encoding can support higher brightness levels. (140 ms exposure time).

show how the solution increases iteration counts regarding defocus blur, color, and image quality. While fewer iterations could provide structurally correct images with incorrect colors, a larger number of iterations help generate color and improve image quality. The provided image quality metrics in Fig. 17 suggest that image quality increases with a larger iteration count, suggesting higher iteration count helps with the replication of features in target images (>500 iterations).

## 15 RESULTS FROM ON-AXIS HOLOGRAPHIC DISPLAY CONFIGURATION

Our methodology is also applicable to holographic display prototypes configured in an on-axis imaging setup. In order to illustrate the adaptability of our approach to on-axis holographic displays, we constructed an additional holographic display configured for on-axis operation.

For the on-axis configured holographic display prototype we decided to employ RGB LEDs as a substitute for lasers. Because of this substitution and the subsequent limited calibration possibilities, our results in this configuration remain as preliminary work. Therefore this section serves the purpose of demonstrating our method can support on-axis configurations. The results obtained from the on-axis configuration are absent from chromatic aberrations, a challange that is faced by off-axis holographic display prototypes. Both DP and direct phase encoded results are presented in Fig. 18 and Fig. 19.

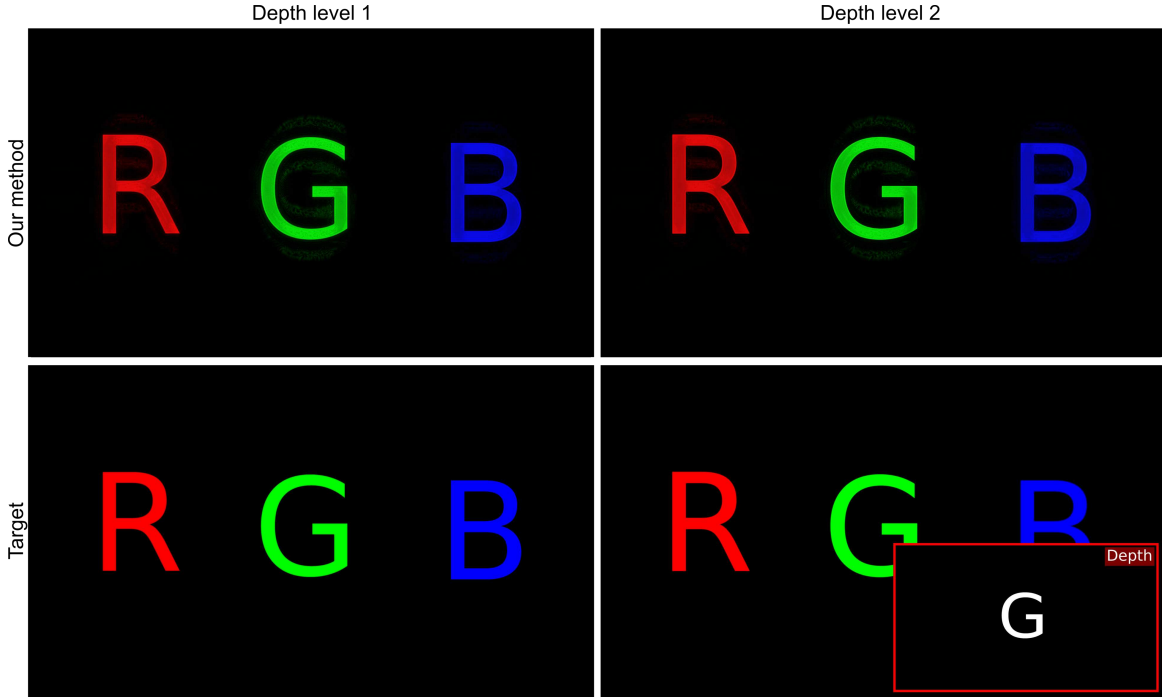

Fig. 12. Simulated reconstruction of our multi-color holograms when they target scenes with spatial color separation (e.g. each letter of a text is dedicated to one color primary) and ×1/8 brightness. Although a corner case, multi-color holograms could not benefit from brightness improvements in such cases while leading to visual distortions.

Here, we provide a list of components used in our on-axis holographic display. It starts with a RGB LED light source, ws2812b (466 nm, 524 nm, 623 nm), that is controlled by an Arduino UNO board. A pinhole aperture, the Thorlabs SM1D12, is positioned infront of the RGB LED source to confine beam divergence. Following to this aperture, we employed a 4f optical setup, composed of the AC508-075-A-ML and AC254-035-A-ML components, thereby further reducing the spot size of our RGB LED light source. Another Thorlabs SM1D12 pinhole aperture follows this 4f system. Positioned after this second aperture is a collimator lens, the AC254-050-A-ML, accompanied by a linear polarizer, the Thorlabs LPVISE100-A. This polarizer is carefully aligned with the fast axis of our phase-only Spatial Light Modulator (SLM) to ensure precise beam alignment. The linearly polarized light beams then reach our phase-only SLM, Holoeye Pluto-VIS. The phase-modulated beam proceeds to a 4f imaging system, consisting of a 75 mm achromatic doublet lens (AC508-075-A-ML), a Thorlabs SM1D12 pinhole aperture, and a 35 mm achromatic doublet lens (AC254-035-A-ML). To capture the resultant image reconstructions, we employ the Ximea MC245CG-SY camera.

## 16 ADDITIONAL DISCUSSIONS

*Eyebox.* Producing a wide eyebox in holographic displays is critical for the success of holographic glasses for Augmented Reality (AR) and Virtual Reality (VR) applications. Recent studies explore eyebox qualities in holographic displays for various optimization methods [11] (e.g. GD or Gerchberg-Saxton). A similar study could help characterize eyebox qualities in our method. In addition, moving away from DP encoding towards direct encoding in our method may help co-optimizing image quality and eyebox size. Currently, our method supports both DP and direct-phase encoding. In our supplementary, we provide early image quality assessments showing noisier

```
1   import torch.optim as optim
2   from odak import propagate_beam,generate_complex_field
3
4   # Provide an initial phase for a hologram (random, manual or learned).
5   φf_n = define_initial_phase(type='random')
6   φf_n.requires_grad = True
7   # Provide number of iterations requested.
8   iter_no= 200
9   # Setup a solver with
10  optimizer = optim.Adam([{'params': φf_n, offset_f_n}],lr=0.002)
11  # Calculate targets for each plane.
12  P_0,P_1,P_2,...,P_n = targetting_scheme(distances)
13
14  # Iterates until iteration number is met.
15  for i in range(iter_no):
16      # Distances between a hologram and target image planes.
17      for distance_id,distance in enumerate(distances):
18          # Clearing gradients.
19          optimizer.zero_grad()
20          # Phase constrain (Equation 5).
21          φ = phase_constrain(φ, offset)
22          # Generates a hologram with the latest phase pattern.
23          O_h = generate_complex_field(1., φ)
24          # Forward model.
25          K = propagate(O_h, distance)
26          # Calculating loss function for the reconstruction.
27          loss += ℒ_m(|U|², P_(distance_id))
28          # Updating the phase pattern using accumulated losses.
29          loss.backward()
30          optimizer.step()
31
32  # Optimized multiplane hologram:
33  φ → O_h
```

Listings 1: Stochastic-Gradient based multiplane phase-only hologram optimization algorithm when reconstructing images at a spatial light modulator plane. The abstraction is Pythonic. Note that this optimization runs for each color channel separately.

results by switching to direct encoding. Our study suggests that regularizing image loss per eyebox size could be necessary for the following works.

*Hardware-in-the-loop.* Our method can also benefit from hardware-in-the-loop techniques [2, 8, 14] to calibrate our holographic display. However, these methods require a dedicated new investigation to operatein multi-color hologram scheme. We leave these for future investigations.

*Diffraction Efficiency.* The fraction of the incident optical power appearing in a diffracted order, diffraction efficiency, is driven by the aperture shape, reflection, and transmission efficiency of an SLM [15]. Regardless of hardware, conventional and multi-color holograms are optimized using diffraction integral simulations. These simulations don't regularize their power distribution beyond the image area. Their hologram-related diffraction efficiency correlates strongly with the design choices of their hologram generation pipelines. We leave a thorough diffraction-efficiency comparison analysis of various pipelines in the literature as a future work.

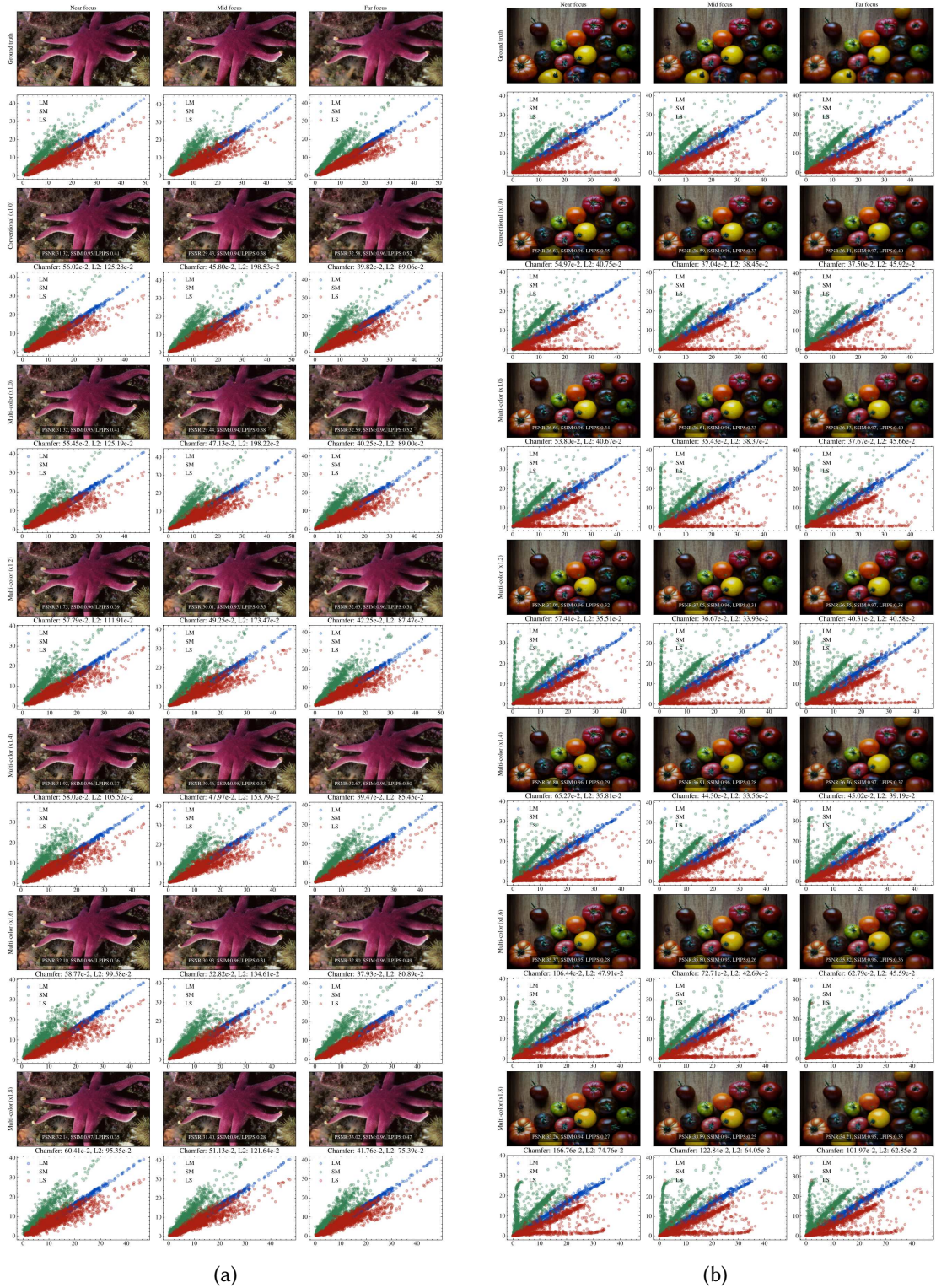

(a)                                                                 (b)

Fig. 13. We evaluate our method's color performance for various target scenes in simulation. We provide a target reconstruction for each case and their perceived color levels in LMS space, precisely plotted LM, LS, and MS pairs (see for exact conversion from RGB to LMS [3]). We observe that our method truthfully generates colors for varying brightness levels in most cases. Sometimes, there could be a perceptible color mismatch when aiming for higher brightness levels (e.g. ×1.8 enhancement).

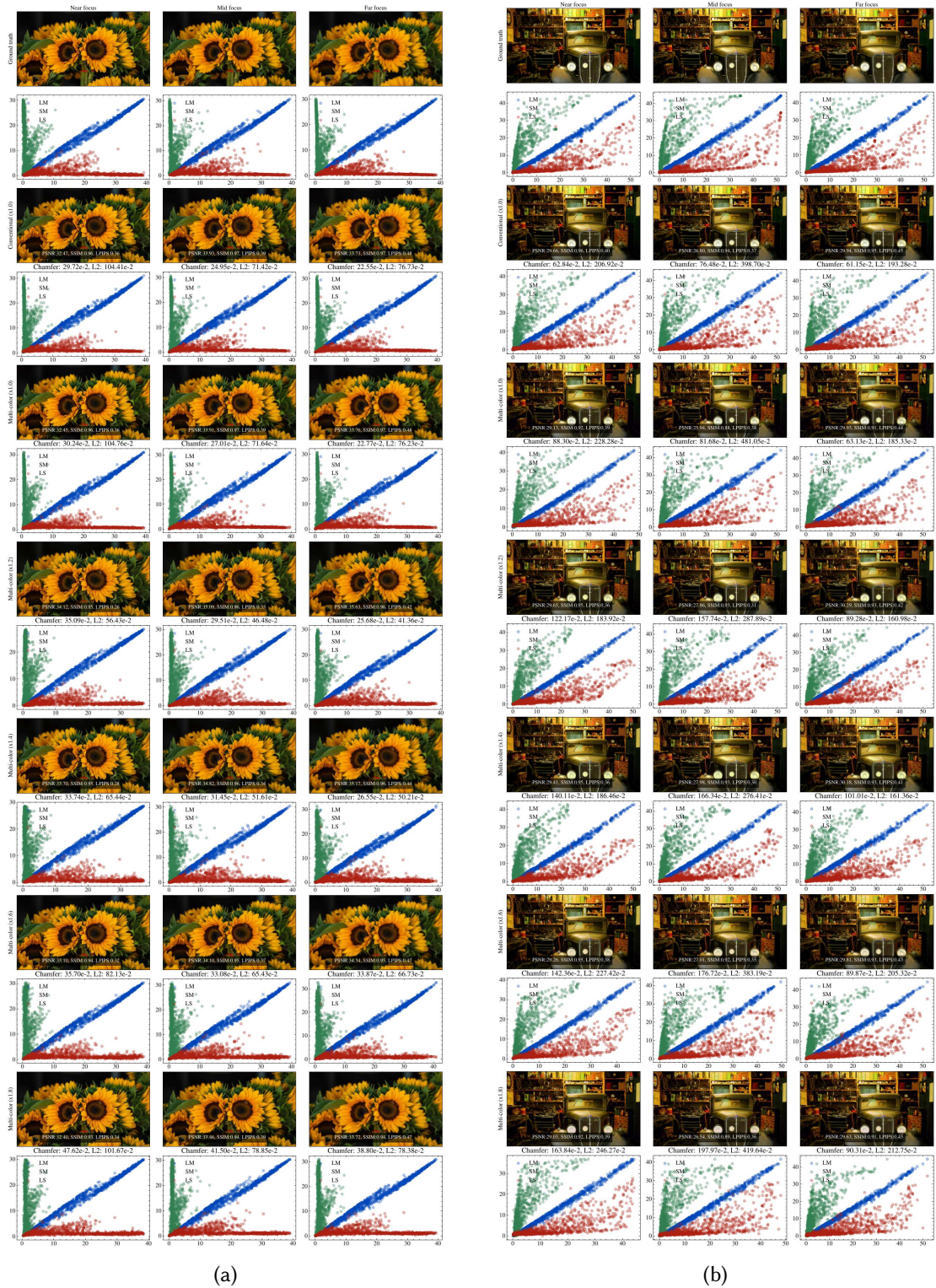

(a)                                                           (b)

Fig. 14. We evaluate our method's color performance for various target scenes in simulation. We provide a target reconstruction for each case and their perceived color levels in LMS space, precisely plotted LM, LS, and MS pairs (see for exact conversion from RGB to LMS [3]). We observe that our method truthfully generates colors for varying brightness levels in most cases. Sometimes, there could be a perceptible color mismatch when aiming for higher brightness levels (e.g. ×1.8 enhancement).

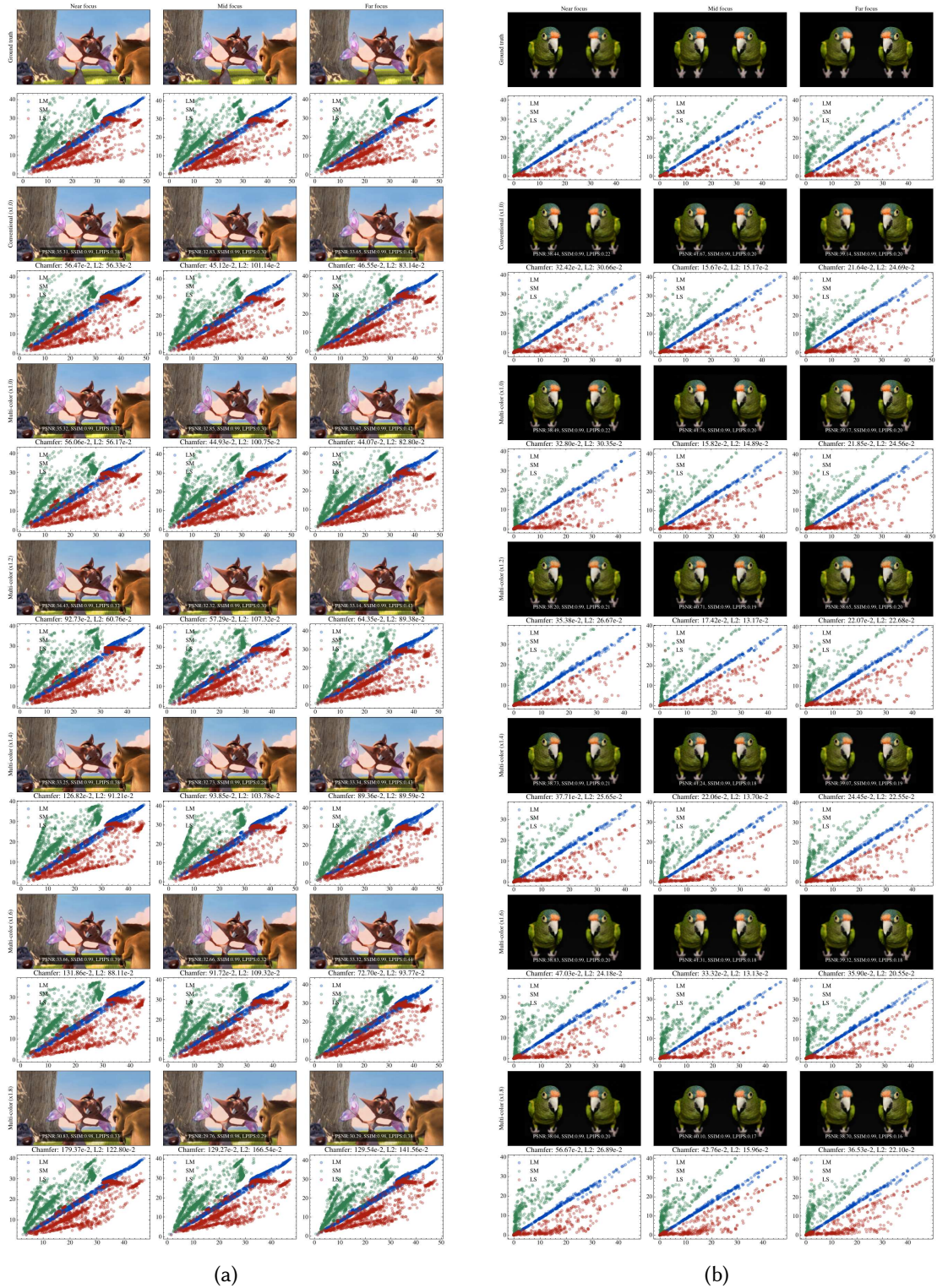

(a)

(b)

Fig. 15. We evaluate our method's color performance for various target scenes in simulation. We provide a target reconstruction for each case and their perceived color levels in LMS space, precisely plotted LM, LS, and MS pairs (see for exact conversion from RGB to LMS [3]). We observe that our method truthfully generates colors for varying brightness levels in most cases. Sometimes, there could be a perceptible color mismatch when aiming for higher brightness levels (e.g. ×1.8 enhancement).

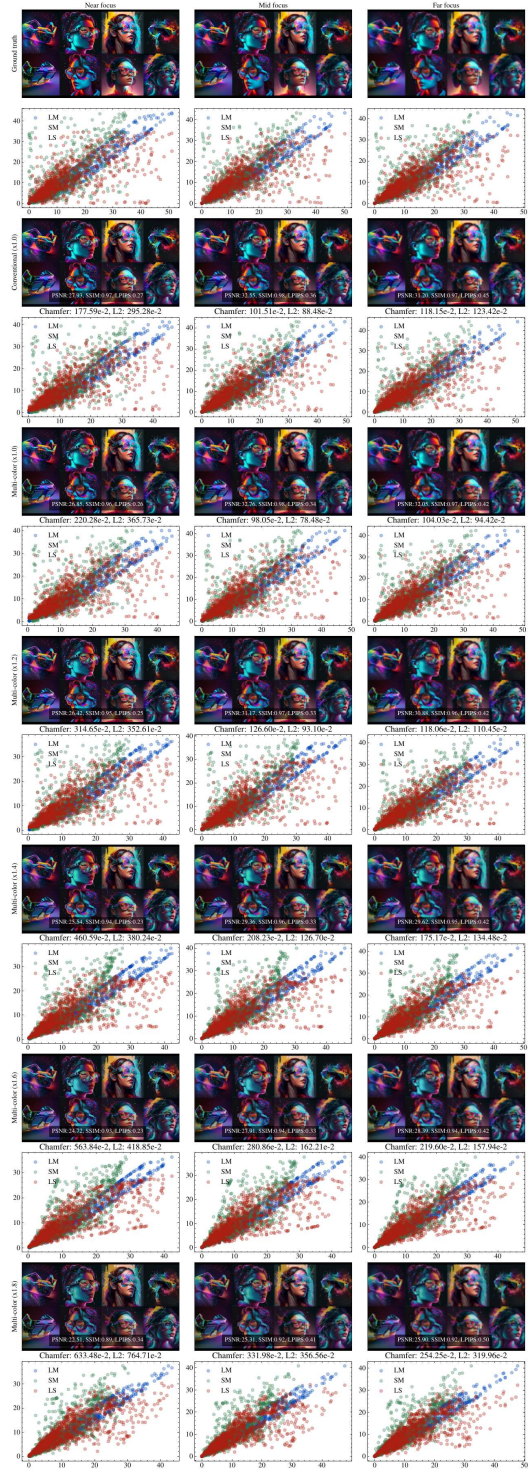

(a)

Fig. 16. We evaluate our method's color performance for various target scenes in simulation. We provide a target reconstruction for each case and their perceived color levels in LMS space, precisely plotted LM, LS, and MS pairs (see for exact conversion from RGB to LMS [3]). We observe that our method truthfully generates colors for varying brightness levels in most cases. Sometimes, there could be a perceptible color mismatch when aiming for higher brightness levels (e.g. ×1.8 enhancement).

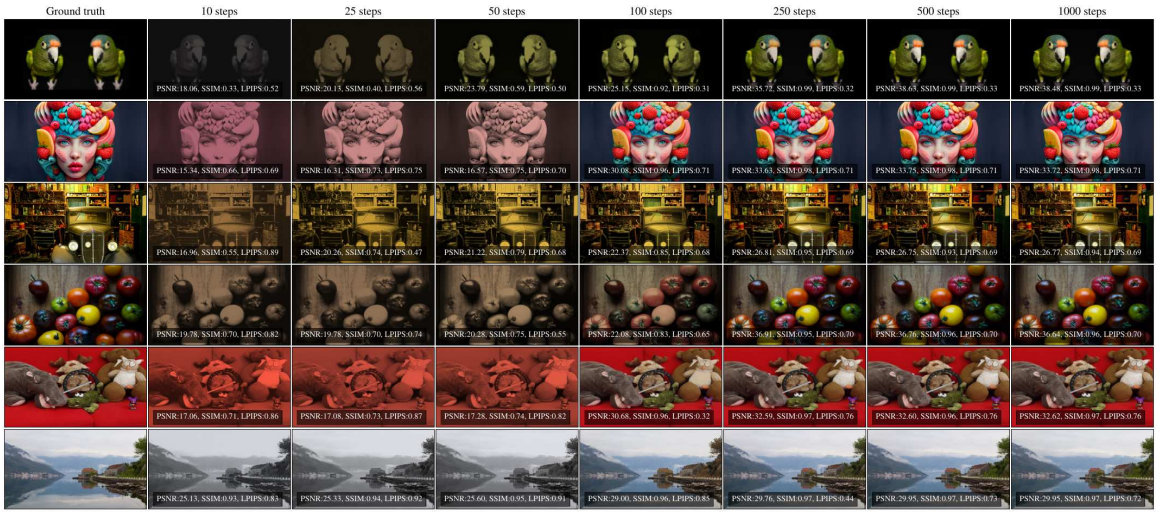

(a)

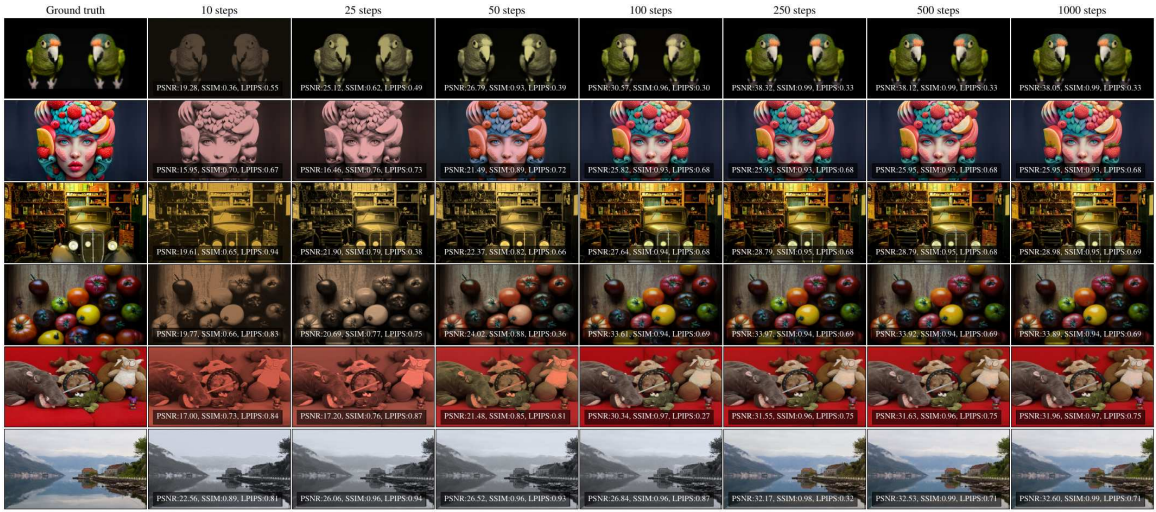

(b)

Fig. 17. We provide two sets of simulated reconstructions for solutions using various iteration count in our method's optimizations. These two sets of simulated reconstructions shows a single depth plane of three plane target scene. The set at the top is generated with a target brightness at ×1.0, while the bottom set is targeting ×1.8.

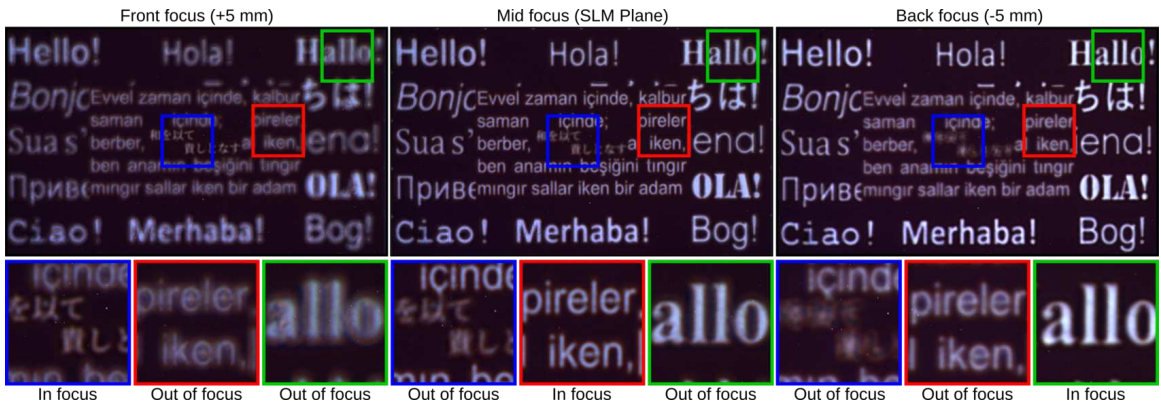

Fig. 18. Three dimensional scene captured with on-axis holographic display prototype using our multi-color DP encoded holograms. Photographs show a multiplane image generated by our multi-color DP coded hologram scheme with three focus planes. The front and back focus planes are remain 5 mm away from the SLM while the middle plane remains on the SLM plane. The targeted brightness level is ×1.8. (450 ms exposure time)

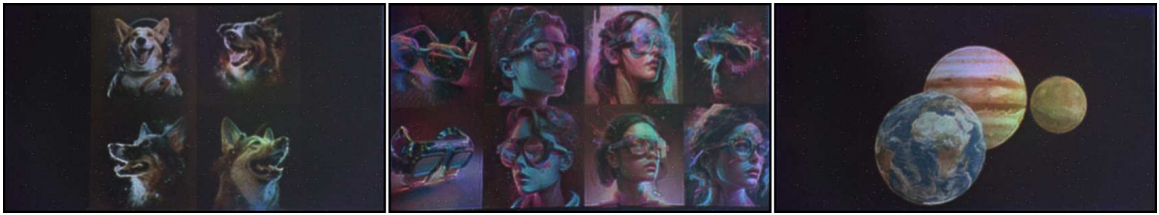

Fig. 19. Images captured with on-axis holographic display prototype using our multi-color direct phase encoded holograms. Photographs show images that are generated 30 mm in front of the SLM using our multi-color direct phase encoded hologram scheme. The targeted brightness levels for all scenes is ×1.8. (450 ms exposure time)

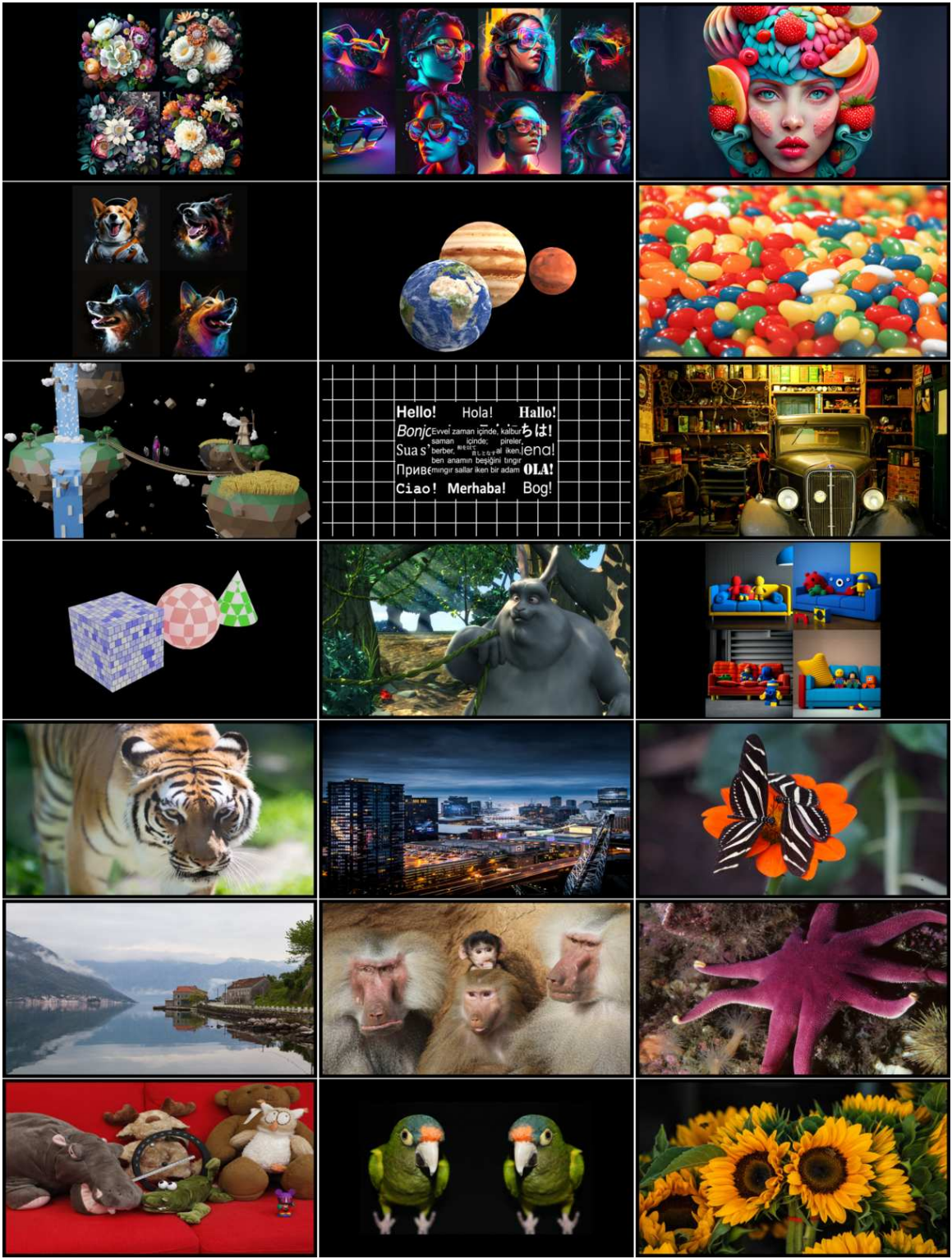

Fig. 20. A gallery of target images used in our manuscript and supplementary (Source link: **Github:complight/images**, DIV2K [1], Image Credits: Planets (self-generated), Candies **Pexels:Voyance Smith**, Floating islands **CGTrader:ShalinSaju123**, Multi-text (self-generated), Primitives (self-generated), Big Buck Bunny **Blender Foundation**, Couch [10], Birds **Unsplash:Ruth Caron**, Other images are created by Midjourney).

}
\end{document}